\documentclass[useAMS,usegraphicx,usenatbib]{mn2e}

\usepackage{ulem}
\usepackage{color}
\usepackage{graphics,graphicx}
\usepackage{times} 
\usepackage{amssymb}
\usepackage{amsmath}
\usepackage{lscape}
\usepackage{url}
\usepackage{multirow}
\usepackage{ctable}
\usepackage{sidecap}
\usepackage{rotating}
\usepackage{threeparttable}
\usepackage{float}
\usepackage{lscape}
\usepackage{tikz}
\usetikzlibrary{snakes}
\newif\ifAMStwofonts
\AMStwofontstrue
\title[ULX Candidates in Nearby Galaxies]{2XMM Ultraluminous X-Ray Source Candidates in Nearby Galaxies}

\author[D.\,J. Walton, T.\,P. Roberts, S. Mateos \& V. Heard]
{D.\,J. Walton$^{1}$ \thanks{E-mail: dwalton@ast.cam.ac.uk}, 
T.\,P. Roberts$^{2}$ \thanks{E-mail: t.p.roberts@durham.ac.uk},
S. Mateos$^{3}$ \&
V. Heard$^{3}$ \\
\footnotesize
$^1$ Institute of Astronomy, Cambridge University, Madingley Road, Cambridge, CB3 0HA \\
$^2$ Department of Physics, Durham University, South Road, Durham DH1 3LE, UK \\
$^3$ Department of Physics and Astronomy, University of Leicester, University Road, Leicester, UK}
\date{}

\def\ro{{\it ROSAT~\/}}

\def\ein{{\it Einstein~\/}}
\def\xmm{{\it XMM-Newton~\/}}

\def\chandra{{\it Chandra}}

\def\epicpn{{\it EPIC}{\rm-pn}}
\def\epicmos1{{\it EPIC}{\rm-MOS1~\/}}
\def\epicmos2{{\it EPIC}{\rm-MOS2 ~\/}}
\def\epicmos{{\it EPIC}{\rm-MOS}}

\def\mpc{\hbox{$\rm\thinspace Mpc$}}

\def\kms{\hbox{$\rm\thinspace km~s^{-1}$}}

\def\H0{{\rm km s$^{-1}$ Mpc$^{-1}$}}

\def\kev{\hbox{$\rm\thinspace keV$}}

\def\ergcms{{\rm ~erg~cm^{-2}~s^{-1}}}

\def\atpcm{{\rm atom~cm$^{-2}$}}

\def\ergps{\hbox{erg~s$^{-1}$}}

\def\msun{\hbox{$\rm\thinspace M_{\odot}$}}

\def\chisq{{$\chi^{2}$}}

\def\rcstat{$C_{\nu}$~\/}

\def\grid25{\hbox{\rm{\small GRID25}}}

\def\etal{et al.~\/}
\def\eg{{\it e.g.~\/}}
\def\etc{{\it etc.}}
\def\ie{{\it i.e.~\/}}

\def\la{\mathrel{\hbox{\rlap{\hbox{\lower4pt\hbox{$\sim$}}}{\raise2pt\hbox{$<$}}}}}
\def\ga{\mathrel{\hbox{\rlap{\hbox{\lower4pt\hbox{$\sim$}}}{\raise2pt\hbox{$>$}}}}}

\def\d25{D$_{25}$}
\def\nh{{$N_{\rm H}$}}

\def\.25{0.25 keV\thinspace}

\def\mbh{\rm $M_{\rm BH}$}

\begin{document}

\pagerange{\pageref{firstpage}--\pageref{lastpage}}
\pubyear{2010}

\maketitle

\label{firstpage}

\begin{abstract}
Ultraluminous X-ray sources (ULXs) are some of the most enigmatic X-ray bright
sources known to date. It is generally accepted that they cannot host black holes
as large as those associated with active galaxies, but they appear to be significantly
more luminous than their better understood Galactic X-ray binary (XRB) cousins, while
displaying an intriguing combination of differences and similarities with them. 
Through studying large, representative samples of these sources we may hope to
enhance our understanding of them. To this end, we derive a large catalogue of
650 X-ray detections of 470 ULX candidates, located in 238 nearby galaxies, by cross
correlating the 2XMM Serendipitous Survey with the Third Reference Catalogue of Bright
Galaxies. The presented dedicated catalogue offers a significant improvement over those previously
published both in terms of number and the contribution of background contaminants,
\eg distant quasars, which we estimate to be at most 24 per cent, but more
likely $\sim$17 per cent. To undertake population studies, we define a `complete' sub-sample of
sources compiled from observations of galaxies with sensitivity limits below $10^{39}$
\ergps. The luminosity function of this sample is consistent with a simple power-law of
form $N(>L_X) \propto L_X^{-0.96 \pm 0.11}$. Although we do not find any statistical
requirement for a cut-off luminosity of $L_c \sim 10^{40}$ \ergps, as has been reported
previously, we are not able to rule out its presence. Also, we find that the number of
ULXs per unit galaxy mass, $S^u$ decreases with increasing galaxy mass for ULXs associated
with spiral galaxies, and is well modelled with a power-law of form $S^u \propto M^{-0.64
\pm 0.07}$. This is in broad agreement with previous results, and is likely to be a
consequence of the decrease in specific star formation and increase in metallicity with
increasing spiral galaxy mass. $S^u$ is consistent with being constant with galaxy
mass for sources associated with elliptical galaxies, implying this older ULX population
traces stellar mass rather than star formation.
\end{abstract}


\section{Introduction}

The first X-ray imaging observations of galaxies beyond the local
group, obtained with the \ein observatory in the early 1980s, lead to
the discovery of a new population of extremely bright, extra-nuclear
X-ray sources with luminosities $L_X > 10^{39}$ \ergps\
(\citealt{Fabbiano89rev}). Such sources are now commonly referred to as
ultraluminous X-ray sources (ULXs), and remain the subject of
continuing speculation as to how they are able to radiate so brightly.
The observed luminosities imply some kind of accreting black hole, but
they are (or at least appear) brighter than the traditional Eddington
limit for the `stellar mass' black holes in X-ray binaries (BHBs)
observed in our own galaxy (\mbh $\sim 10$\msun). However, it is
generally accepted that these are not extra-nuclear examples of the
supermassive black holes (SMBH; \mbh$~>10^{5-6}$\msun) expected to
power active galactic nuclei, as such objects should sink to their
galaxy centres comfortably within a Hubble time due to dynamical
friction (\citealt{MilCol04}).

Improvements in the available X-ray instrumentation, especially with
the launch of the \xmm and \chandra\ observatories, combined with
multi-wavelength studies have improved our understanding of ULXs
considerably in recent years, but the exact cause of the apparently
extreme luminosities has yet to be determined unambiguously for any
individual source, let alone for the population as a whole.  Most of
the proposed explanations revolve around one of the three following
ideas. The first is that the black holes powering ULXs are simply much
more massive than those in BHBs (see \eg \citealt{Colbert99}), perhaps
the long sought after intermediate-mass black holes (IMBHs;
$10^{2}$\msun$ \lesssim$ \mbh\ $\lesssim 10^{4}$\msun), accreting at
rates substantially below the Eddington limit as per Galactic BHBs.
The second is that the black holes hosted by ULXs are essentially the
same as those observed in stellar mass BHBs, but in an accretion state
in which they are able to radiate at or even above the Eddington limit
(see \citealt{Remillard06rev} for a recent review of the standard BHB
accretion states). A number of methods by which this may be physically
possible have been suggested (see \eg \citealt{Poutanen07};
\citealt{Finke07}; \citealt{Abram80}).  Indeed, observational evidence
from ULXs now appears to show differences between ULXs and Galactic
BHBs, suggestive of ULXs representing a new super-Eddington accretion
state (\citealt{Roberts07}; \citealt{Gladstone09}).

The third idea is that ULXs do not emit isotropically, and are sources
that we view at a favourable orientation (\citealt{King01}). Depending
on the extent to which the emission is anisotropic, this can
potentially lead to artificially inflated estimates for
the luminosity of a source, which is usually calculated under the
assumption of isotropy, and again there may be no need to invoke
larger black holes than those present in stellar mass BHBs.  In the
context of ULXs, it has been suggested that the required anisotropic
emission may be attained through \eg geometric collimation of
radiation due to a thick accretion disc (possibly the `slim' discs
proposed to allow super-Eddington emission, see \citealt{Abram80}), or
through collimation into relativistic jets similar to those observed
in other accreting sources (\citealt{Reynolds97}). However, a growing
number of ULXs are observed to be embedded within roughly spherically
symmetric photoionised nebulae (\citealt{Pakull03}, \citealt{Kaaret04}).
Detailed study of these nebulae suggests they are not supernovae remnants
(SNRs) but are apparently inflated by the ULX (\citealt{Pakull08}).
Strongly anisotropic emission may be ruled out in these cases through
morphology and photon counting arguments, so it appears unlikely ULXs as
a class may be considered `microblazars'. 

We are now in the position where recent scenarios explaining the
physical processes governing the high luminosities of ULXs combine two
or more of the ideas presented above.  For example, super-Eddington
discs are predicted to inflate and produce outflowing winds that act
to collimate the X-ray emission from their central regions
(\citealt{Begelman06}; \citealt{Poutanen07}; \citealt{King08}).  Also,
it has been suggested that a proportion of the ULX population may
consist of larger stellar remnant black holes (up to $\sim 80$\msun),
that may plausibly be formed in low-metallicity environments,
accreting at mildly super-Eddington rates (\citealt{Zampieri09}).

As previously stated, unambiguously distinguishing between these
explanations, notably the IMBH and stellar mass interpretations, has
proven extremely difficult on the basis of current observations. For a
recent review on this topic, see \cite{Roberts07}. In large part, this
is because reliable dynamical mass estimates are not yet available for
any ULXs.  Until such mass estimates become reality, an excellent way
in which to enhance our understanding of ULXs is through population
studies of large ULX samples, as compiled by \cite{Colbert02},
\cite{Swartz04}, \cite{LiuBregman05} and \cite{LiuMirabel05}. For example,
such work has demonstrated that ULXs are strongly linked with recent
star formation and are prevalently located in spiral galaxies (see
\eg \citealt{Swartz04}; \citealt{Grimm03} and \citealt{Gilfanov04a};
\citealt{Liu06}; but especially \citealt{Swartz09} and references therein).
This relationship with star formation strongly suggests that the majority
of ULXs are a form of high mass X-ray binary (HMXB), where a
massive young star forms the necessary large mass reservoir for the
ULX (see \eg \citealt{Rappaport05}). It is also suggested that most of
the ULXs observed in elliptical galaxies are likely to be low mass
X-ray binaries (LMXBs; \citealt{Colbert04},
\citealt{Humphrey03}). Based on the luminosity function of a very
large sample of HMXBs, and the relation between the star formation
rate and the total X-ray luminosity of star forming galaxies,
\cite{Grimm03} and \cite{Gilfanov04b} argue that there may be an upper
limit of $\sim$100\msun\ to the black holes in standard HMXBs, which
if true has important consequences for ULXs.

In addition, the compilation of large samples of ULXs can identify
extreme examples of these sources.  This is pertinent as it has become
apparent that there exists a rare subgroup of the ULX population which
display even more extreme X-ray luminosities of $L_X > 10^{41}$
\ergps. These have been dubbed Hyper-Luminous X-ray Sources (HLXs;
\citealt{Gao03}), and may be the most promising candidates in terms of
the search for IMBHs. To date the most notable of these is ESO 243-49
HLX-1, observed to reach luminosities of $1.2\times10^{42}$ \ergps\
(\citealt{Farrell09}), but only a handful of other candidate HLXs are
known (\eg \citealt{Davis04}; \citealt{Jonker10}).  Clearly finding
more such objects is an interesting and potentially illuminating
objective.

Here, we embark upon such a study and present a large catalogue of
X-ray detections of candidate ULXs, compiled from Data Release 1 of
the 2XMM Serendipitous survey (2XMM-DR1, hereafter referred to as
2XMM). The paper is structured as follows: section 2 details the
observations used and the process of compiling the catalogue, section
3 presents and considers some of the science that may be derived from
it, and finally section 4 summarises our conclusions.

\section{Catalogue Production}

As stated, the major resource utilised for our source selection was
the 2XMM-DR1 catalogue. This is described by \cite{2XMM}. In brief, this
is a catalogue of 246897 detections drawn from 3491 public \xmm fields
imaged by the European Photon Imaging Camera (EPIC) detectors, data
spanning the first 7 years of the mission. It contains 191870 unique
X-ray sources (as some sources have been multiply imaged, given that
\xmm has studied some fields on several occasions), and covers roughly
$\sim 1\%$ of the whole sky. Such a large catalogue presents a
hitherto unrivalled opportunity to study relatively rare source
classes such as ULXs; in fact, at the modal detection flux for a 2XMM
source ($\sim 2 \times 10^{-14} \ergcms$ in the 0.2--12.0\,keV band;
see Fig. 9 of \citealt{2XMM}) one can roughly expect to detect all
isolated ULXs within the catalogue field-of-view out to $d \sim 20$ Mpc.
We therefore took the entire 2XMM catalogue as the starting point for
our ULX candidate selection. This section describes the process of
condensing 2XMM into a catalogue of ULX candidates, and the properties
of the derived catalogue.

\begin{figure*}
\centering
\includegraphics[width=8.5cm,angle=270]{figs/n4472xmm.ps}\hspace*{0.5cm}
\includegraphics[width=8.5cm,angle=270]{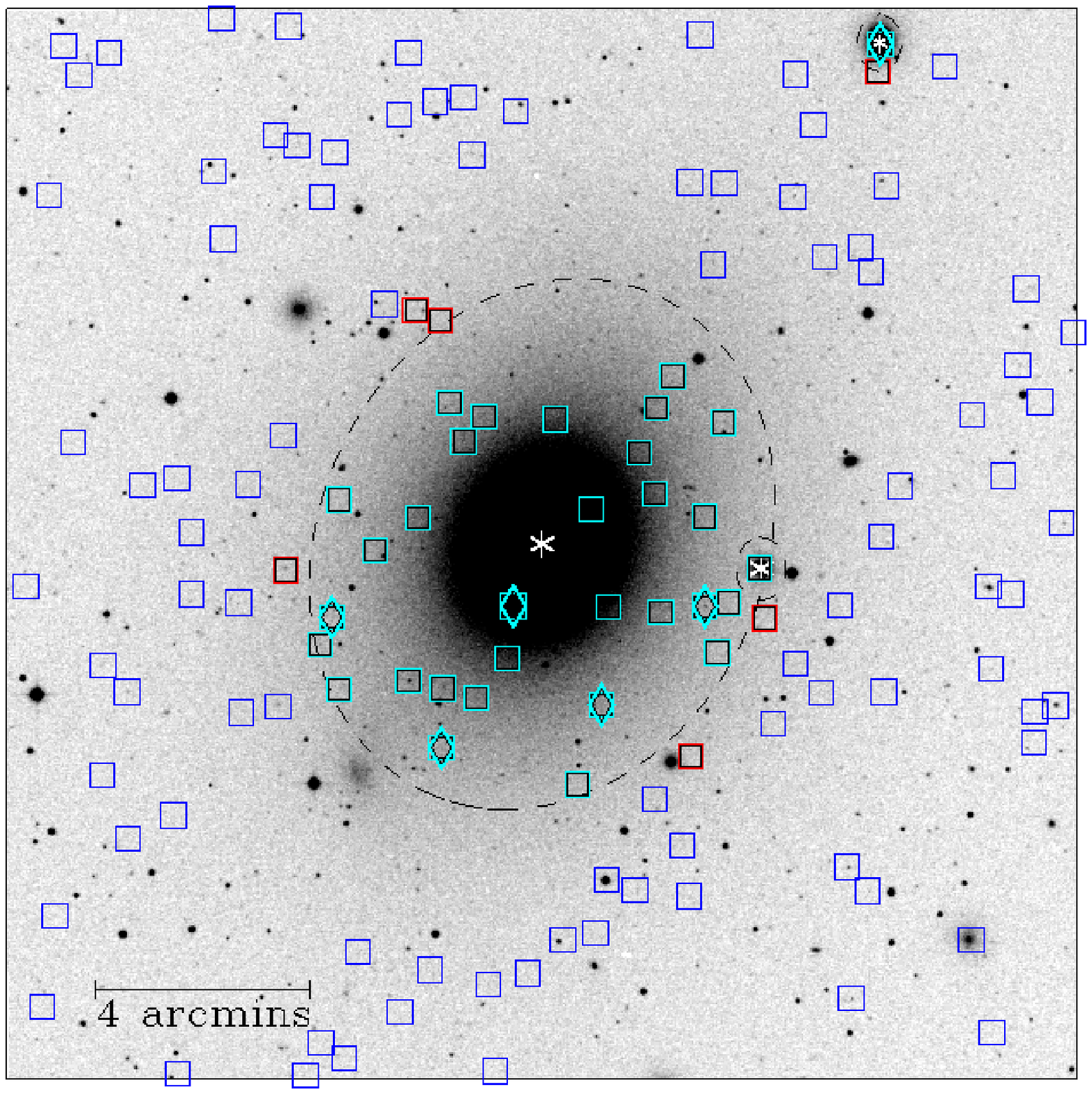}\vspace*{0.5cm}
\includegraphics[width=8.5cm,angle=270]{figs/n5194xmm.ps}\hspace*{0.5cm}
\includegraphics[width=8.5cm,angle=270]{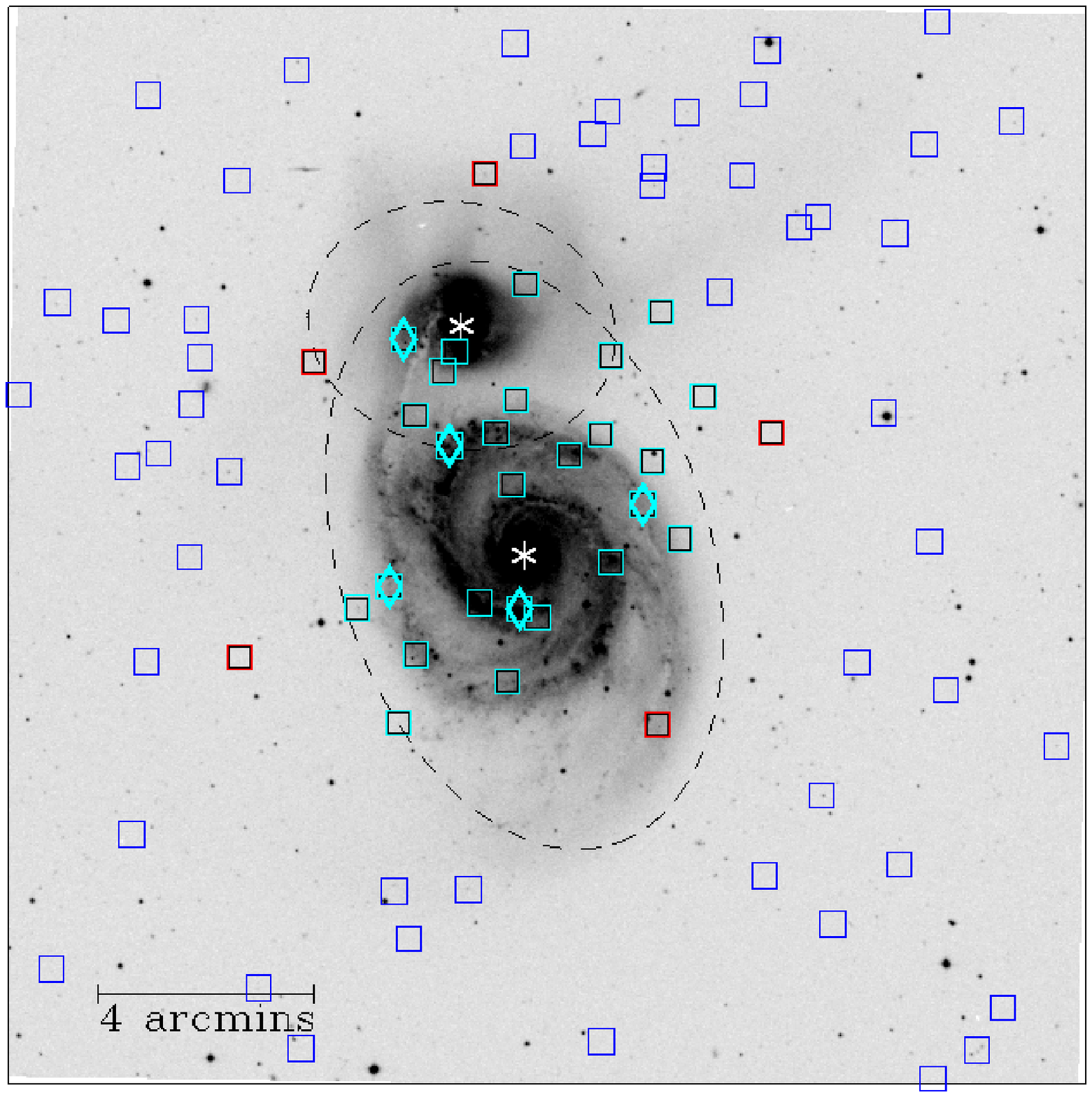}
\caption{An illustration of the 2XMM data and ULX selection process.
We show data for two galaxy fields: ({\it upper panels\/}) NGC 4472,
also including NGC 4464 (to the north-west of NGC 4472, where north is
up in all panels) and NGC 4467 (west of NGC 4472, overlapping its \d25
ellipse); ({\it lower panels\/}) NGC 5194, overlapping with NGC 5195
to its north.  In all panels the \d25 isophotal ellipse is shown, and
the galaxy nuclear position is marked with an asterisk.  ({\it left
panels\/}) A combined EPIC data broad-band (0.2--12.0\,keV) image with
source detections overlaid.  The images were extracted as pipeline
products from the \xmm science archive (XSA), and are convolved with a
1-pixel ($\equiv 4$ arcsec) HWHM Gaussian mask for display purposes.
The images are displayed on a logarithmic heat scale, between
2.5-100000 count pixel$^{-1}$ (NGC 4472), and 1-10000 count
pixel$^{-1}$ (NGC 5194).  Point source detections are marked by cyan
open squares, and extended source detections by green circles with
radius equivalent to the measured radius (note that this is capped at
80 arcsecs in the 2XMM catalogue).  ({\it right panels\/}) Digitised
sky survey (version 2) red images of the galaxy fields, with point
source positions overlaid.  Negative images, linearly scaled, are
shown for display purposes.  The catalogue creation process is
highlighted by showing field sources as blue open squares, with
sources within the \d25 great circle but not the \d25 isophotal
ellipse shown by (thicker) red squares.  Sources within the ellipses
are shown as cyan open squares with black inner lining, with those
selected at ULX-like luminosities highlighted by an additional diamond
symbol.}
\label{gals}
\end{figure*}

\subsection{Initial Correlations and Filtering}

The initial selection of ULX candidates was performed by
cross-correlating the 2XMM catalogue with the Third Reference
Catalogue of Bright Galaxies (RC3; \citealt{RC3}). This catalogue
contains 23022 individual galaxies, the majority of which have either
isophotal diameters $D_{25} > 1$ arcminute, total $B$-band magnitude
$B_{\rm T} < 15.5$ or recessional velocity $cz < 15000$\kms.  Before
cross-correlation with 2XMM, galactic centre positions were updated
and missing recession velocities obtained via a cross-correlation of
RC3 with the online NASA Extragalactic Database
(NED)\footnote{http://nedwww.ipac.caltech.edu/} and the VIZIER/SIMBAD
databases\footnote{http://cds.u-strasbg.fr/}. Galaxies that remained
without a recession velocity after these measures were excluded from
our analysis. Distances to galaxies with recession velocities less
than 1000\kms, for which peculiar motions may dominate the Hubble
flow, were obtained by cross-correlation with the nearby galaxy
catalogue of \cite{TULLY}. For the other galaxies, the Hubble flow is
considered a good approximation, and distances were obtained using a
Hubble constant of $H_{0} = 75$\,\H0 to remain consistent with
\cite{TULLY}.

Cross-correlation of RC3 with 2XMM was performed using the
\textsc{topcat} graphical user interface\footnote{Tool for Operations
on Catalogues and Tables; http://www.star.bris.ac.uk/$\sim$mbt/topcat/}.
An initial match was performed between 2XMM and RC3 via a conical
search, within a radius equal to half of the RC3 galactic \d25
semi-major axis (where the \d25 ellipse is defined as the elliptical
equivalent to the 25th magnitude isophote for a galaxy). An elliptical
filter was then applied to the matched source list to select only those
sources that lay within the \d25 elliptical isophote of all the
galaxies, based on the RC3 major and minor axis ratios and position
angles. However, a number of galaxies included in RC3 do not have
position angles listed. In these cases, we instead performed a
circular match within the minor axis. At this point we also discarded
all sources flagged as extended, as ULXs should by definition be
point-like (although a more detailed discussion of this exclusion is
given in section \ref{sec_lim}). The success of this process was
verified by visual inspection for a number of galaxies; we show the
examples of NGC 4472 and NGC 5194/5 in Fig. \ref{gals} for illustration,
as well as to highlight the complexities and limitations of the 2XMM
source selection process.

X-ray luminosities were calculated for all remaining sources using the
full band EPIC flux (0.2--12.0\,\kev), as given in the 2XMM catalogue,
and the distances to the putative host galaxies obtained earlier.
The 2XMM fluxes are calculated separately for each of the three EPIC
detectors from the observed count rates, assuming an absorbed powerlaw
model for the average source spectrum ($\Gamma$ = 1.7, \nh = $3 \times
10^{20}$\,\atpcm). The quoted EPIC flux is then the weighted average of
the fluxes obtained for the individual detectors (\citealt{2XMM}).
Although luminosity uncertainties are likely to be dominated by the
uncertainty in the distance to the galaxies, particularly where local
peculiar motions may be important, the uncertainties in these distances
are not well-quantified (although any error based on our choice of
$H_{0}$ will at least be systematically propagated through the sample).
We therefore calculated the uncertainty on the luminosity solely from
the uncertainty on the flux. As dictated by the standard definition of
a ULX, sources with $L_{X} < 10^{39}$ \ergps\ were discarded, although
those that were consistent within 1$\sigma$ of such luminosities were
retained in order to include the largest number of possible ULX
candidates.  We also adopted a slightly conservative stance on the
quality of detection to retain, requiring them to be at least
3.5$\sigma$ detections according to 2XMM. This analysis resulted in a
large preliminary catalogue of source detections at (or above) ULX
luminosities, and a small sample of slightly fainter X-ray sources.

\subsection{Filtering for Known Contaminants}
\label{sec_clean}

\begin{table*}
\centering
\caption{Some brief details of the detections included in, and the host
galaxies contributing to, the presented catalogue and some of its key
subsets. Notes: both the average X-ray flux $\tilde{f}_{\rm X}$, and the
average total counts per source (summed across all three EPIC detectors)
$\tilde{C}$, are given as the median value across all ULX candidate
detections.  All other averages are the arithmetic mean, and are shown
with the calculated error on this value. The average X-ray flux and
luminosity values are for the full 0.2--12.0 keV energy band.
}
\begin{tabular}{lcccc}\hline
	& Full catalogue	& Spiral sample	& Elliptical sample 	& Complete sample \\\hline
Total detections						& 650	& 450	& 200	& 242 \\
Individual candidate ULXs				& 470	& 305	& 165	& 169 \\
\hspace{2mm}(with multiple detections		& 101	& 70		& 31		& 31) \\
Host galaxies							& 238	& 141	& 97		& 71 \\
\hspace{2mm}(with multiple candidate ULXs	& 113	& 79		& 34		& 43) \\
 & & \\
& \multicolumn{4}{c}{{\it per host galaxy}} \\
$\langle d \rangle$ (Mpc)					& $37.9 \pm 2.2$	& $29.9 \pm 2.1$	& $49.6 \pm 4.0$	& $13.1 \pm 0.7$ \\
$\langle M_{\rm B} \rangle$				& $-20.5 \pm 0.1$	& $-20.3 \pm 0.1$	& $-20.7 \pm 0.1$	& $-20.2 \pm 0.1$ \\
 & & \\
& \multicolumn{4}{c}{{\it per ULX candidate detection}} \\
$\tilde{f}_{\rm X}$ ($\ergcms$)				& $4.2 \times 10^{-14}$	& $4.8 \times 10^{-14}$	& $3.3 \times 10^{-14}$	& $1.4 \times 10^{-13}$ \\
$\tilde{C}$ (ct)							& 293	& 314	& 279	& 1160 \\
$\langle$ log $L_{\rm X} \rangle$ ($\ergps$)	& $39.48 \pm 0.02$	& $39.46 \pm 0.02$	& $39.54 \pm 0.04$	& $39.31 \pm 0.03$ \\
$\langle$ HR1 $\rangle$					& $0.48 \pm 0.01$	& $0.52 \pm 0.02$ & $0.41 \pm 0.03$	& $0.55 \pm 0.02$ \\ 
$\langle$ HR2 $\rangle$					& $0.23 \pm 0.01$	& $0.26 \pm 0.02$ & $0.16 \pm 0.03$	& $0.30 \pm 0.02$ \\ 
$\langle$ HR3 $\rangle$					& $-0.27 \pm 0.01$	& $-0.23 \pm 0.02$ & $-0.36 \pm 0.02$	& $-0.22 \pm 0.02$ \\ 
$\langle$ HR4 $\rangle$					& $-0.46 \pm 0.01$	& $-0.46 \pm 0.02$ & $-0.45 \pm 0.03$	& $-0.50 \pm 0.01$ \\ \hline
\end{tabular}
\begin{minipage}[t]{3.3in}
\end{minipage}
\label{catsum}
\end{table*}

The largest identifiable contaminant of our sample of ULXs were the
active galactic nuclei (AGNs) within the ULX host galaxies. A simple
cut to remove classic AGNs at $L_{X} \geq 10^{42}$ \ergps\ would
however be insufficient for our purposes as it would not remove the
potential low-luminosity AGNs (LLAGNs) contaminating our
sample. Unfortunately, LLAGNs overlap significantly with ULXs in terms
of observed luminosity, with LLAGN luminosities as low as $\sim
10^{38}$ \ergps\ (\citealt{Ghosh08}; \citealt{Zhang09}), hence making
ULXs and LLAGNs indistinguishable by X-ray luminosity alone. The only
way to circumvent this problem is to remove all possible LLAGN
candidates; hence we were required to remove all sources in close
proximity to the nucleus.

\begin{figure}
\centering
\includegraphics[width=5.6cm,angle=270]{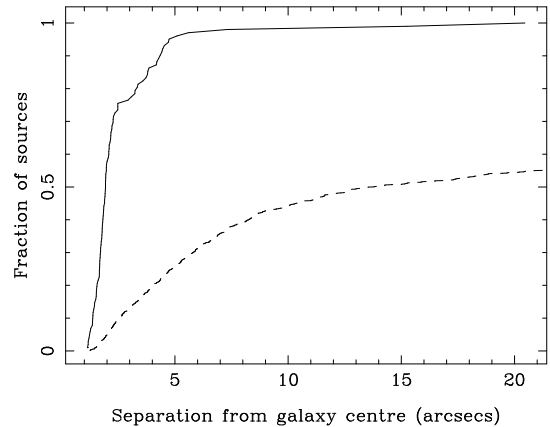}
\caption{Cumulative distribution plot, showing the separation of the
filtered 2XMM detections from the centres of the host galaxies.  We
plot this only for sources in close proximity to the nucleus ($r_{\rm
max} \la 20$ arcsecs), and show the sample of classic AGNs (solid
line), and the comparison ULX/LLAGN sample (dashed line) distributions
separately.}
\label{agns}
\end{figure}

They key question then is: what exclusion radius around the centre of
the galaxy will optimise the removal of LLAGN candidates, while
retaining the maximum number of ULXs? We approached this in an
empirical fashion by defining a conservative maximum separation of
each source from the nuclear position, $r_{\rm max}$ as the sum of its
calculated separation and the $3\sigma$ error in its position as
defined in 2XMM. We then selected the classic AGN contamination of the
sample by discarding any sources with $L_{\rm X} < 10^{42}$ \ergps,
leaving 102 likely {\it bona fide\/} AGNs.  We plot the separation of
these sources (quantified as the maximum separation, $r_{\rm max}$)
from their host galaxy central position as a cumulative distribution
in Fig. \ref{agns}.  The same distribution for all sources with
$10^{39} < L_{\rm X} < 10^{42}$ \ergps, taken from the same working
version of the catalogue, is also shown in Fig.~\ref{agns} in order to
illustrate how any empirical cut would affect this section of the
dataset.

The initial cut was taken at 5'', as this removed 95\% of the AGN
sample, but only 25\% of the sources with luminosities in the ULX
range.  As a significant fraction of the latter sample are likely to
be LLAGNs, the number of actual ULXs discarded is unlikely to have been
too large. However, an inspection of the remaining catalogue revealed
that there were still a significant number of known and probable AGNs
present.  We therefore tried larger excision radii, and found that a
value of $r_{\rm max} = 7.5$'' struck the best balance between
removing known AGNs and leaving a large sample of ULX candidates ($<
40\%$ of the ULX-luminosity sample are removed, many of which we
suspect to be LLAGNs at these small nuclear separations).

The remaining sample was further cleaned with the following steps.
Firstly, all high luminosity sources ($L_X > 10^{40} $ \ergps) were
inspected by checking their host galaxy nuclear classification, their
separation from the galaxy centre, and the size of their position
error. We identified further examples of AGNs that had not been
filtered by the excision radius due either to a poorly-defined nuclear
position for the galaxy, or to a large position error, with the latter
resulting from the source being weakly detected and/or well off-axis
and/or in a region with a high diffuse background component; these
were also discarded. There were also a number of high luminosity
sources close to (within $\sim$\,15'' of) the centres of elliptical
galaxies that were not classified as AGNs. A conservative stance was
also taken with these objects, many of which were poorly parameterised
due to the presence of a strong diffuse ISM in the galaxies, as we
were concerned the 2XMM catalogue may not be differentiating between
peaks in the strong ISM and true point sources, hence these sources
were also excluded. The second step related to sources that were
multiply-detected in 2XMM, where only some detections lay within the
$r_{\rm max} = 7.5$ arcsec excision radius (which again was mainly due
to poorer position constraints on some detections). All detections of
these sources were excluded from the catalogue.

These measures remove the contamination of the sample by AGNs
associated with the host galaxies. However, other AGNs may still be
present in the field of view, predominantly at cosmic distances behind
the host galaxies, some of which will be known objects. We therefore
cross-correlated our remaining catalogue with the NED and
SIMBAD/VIZIER databases to search for known QSOs and other potential
contaminants.  We found a total of 54 detections of 39 suspected
contaminants coincident with our host galaxies, including 17 AGNs, 7
foreground stars and 13 possible X-ray detections of recent supernovae
within the host galaxies. All these detections were discarded from our
catalogue. Of course, this analysis will not correct for any
previously {\it unknown\/} contamination of the sample with background
or foreground objects, the magnitude of which was relatively large in
previous ULX samples. We return to this problem in \S
\ref{sec_bg_contam}.

Finally, there were a number of other minor issues that required
attention during the production of the cleaned catalogue. Close
inspection revealed a number of repeated entries of the same detection
of individual sources. This occurred where the source position lay
under the \d25 ellipse of multiple galaxies (see Fig~\ref{gals} for
some examples). In this situation we assumed the source was truly
associated with the galaxy whose centre it was closest to, and removed
all other entries of the same detection. In addition, we also
inspected the combined EPIC images of detections with small
separations from their respective galactic centres to remove
detections within the pattern of scattered light produced by the
mirror supports for bright sources, which are likely to be artificial.

Once we had our catalogue of good ULX candidates, we returned to 2XMM
to re-append any detections of these sources that had previously
been discarded. This allowed for sources with variable luminosity, sources
located right at the edge of the \d25 isophote of their respective galaxies
with detection positions scattered on either side, \etc\ It is after
this analysis and filtering procedure that we present the catalogue.

\subsection{The Catalogue}
\label{sec_cat}

\begin{figure*}
\begin{center}
\rotatebox{0}{
{\includegraphics[width=475pt]{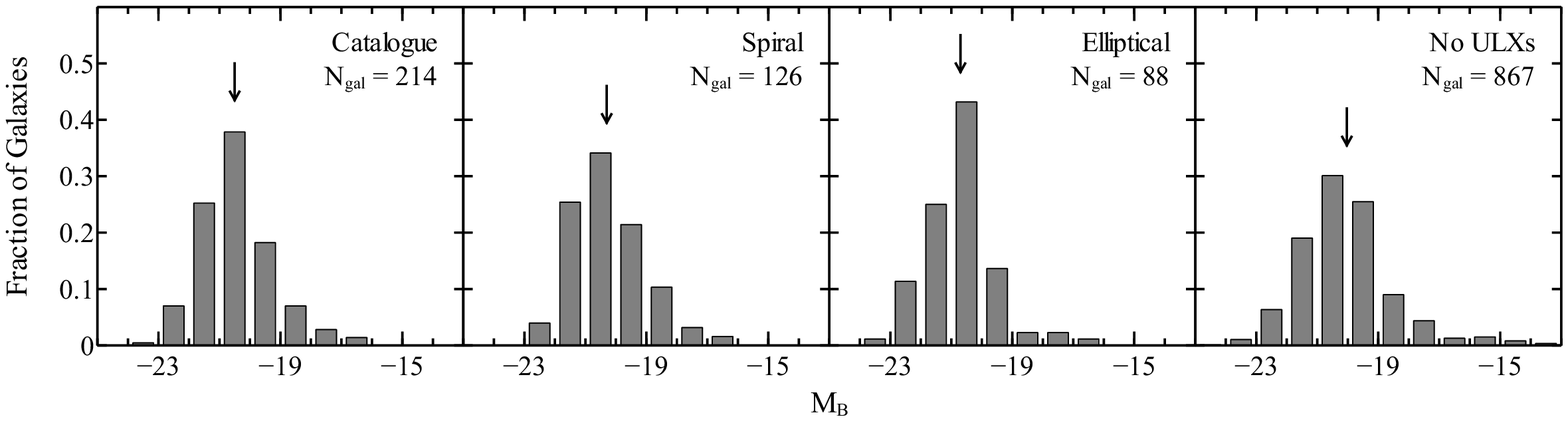}}
}
\rotatebox{0}{
{\includegraphics[width=475pt]{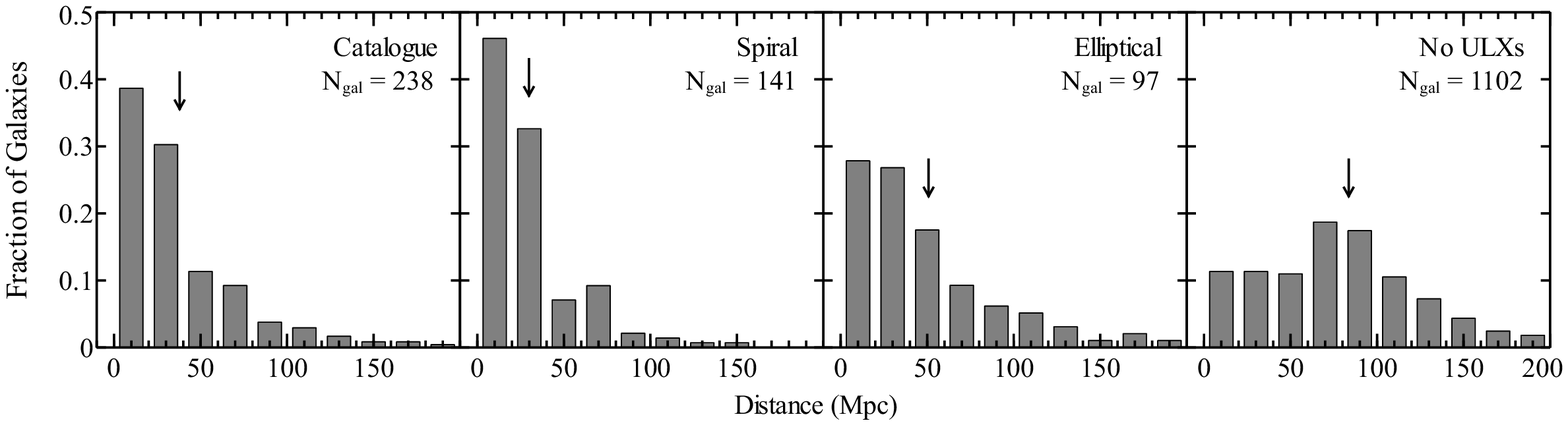}}
}
\end{center}
\caption{Fractional distributions for the absolute B band magnitude
(magnitude bins) and the distance (20 Mpc bins) for RC3 galaxies
that host ULX candidates. Arrows indicate the values quoted in Table
\ref{catsum}. Elliptical galaxies are in general brighter, and more
distant than their spiral galaxy counterparts. For comparison, we
also show the distributions of the RC3 galaxies covered by 2XMM-DR1
observations that did not contribute any ULX candidates to our
sample. Note that the gaps in the histograms are for aesthetic
purposes only, and that not all RC3 galaxies have B-band magnitude
entries.}
\label{fig_gal_hist}
\end{figure*}

The filtering processes left us with a catalogue comprising of 650
detections of 470 individual candidate ULXs, 101 of which were
detected on two or more occasions. Of the detections included, 121
have $L_{X} < 10^{39}$\ergps, of which 79 are within 1$\sigma$ of
the classic ULX luminosity limit; the rest are detections of sources
with variable luminosity which have been observed to emit at $L_{X}
> 10^{39}$\ergps\ at other epochs. The catalogue contains 369 data
entries per candidate ULX detection, including the full RC3 entry
for the host galaxy, the full 2XMM entry for the X-ray source, and
additional information we have added in the derivation of the
catalogue (distance, X-ray luminosity, \etc). For ease of comparison,
we highlight probable common entries between our catalogue and those
of \cite{Colbert02}, \cite{Swartz04}, \cite{LiuBregman05} and
\cite{LiuMirabel05}. The catalogue contains 367 sources that do
not appear to be in any of those works, that we flag as `new' ULX
candidates. The position errors on the detections are on average only
1'', and are always less than 3.5''. Additional summary details of
the catalogue are presented in Table \ref{catsum}, as well as a
detailed example of some catalogue entries in Appendix A. The full
catalogue itself is available electronically with this
paper\footnote{Catalogue to be made available online via MNRAS;
available in the meantime via private communication}.

In this section, and much of the rest of this paper, we choose to
treat the ULX candidates hosted by spiral galaxies and those hosted by
elliptical galaxies separately. There are very clear reasons for
making this separation. Previous work has shown that the X-ray source
populations of galaxies are dependent on two factors: the mass of the
galaxy, which relates to an underlying population of older LMXBs, and
the amount of ongoing star formation, which relates to the HMXB
population (e.g. \citealt{Colbert04}; \citealt{Humphrey03};
\citealt{Lehmer10}). In terms of ULXs, it is clear that many of the
ULXs in spiral galaxies - particularly in hosts with a high star
formation rate - are directly related to the star formation and are
likely a type of high-mass X-ray binary system (see \citealt{Swartz09}
and references therein).  This large-scale star formation is not
present in elliptical galaxies, and so a separate, physically distinct
type of ULX must be present in these systems, as argued by \cite{King02}.
We therefore distinguish elliptical and spiral candidate ULXs in our
analysis, making the distinction based on the numerical index for the
Hubble type, $T$, as registered in RC3. Candidate ULXs in types SO/a and
later ($T \geq 1$) are taken to comprise the spiral sample (which
therefore includes all morphologically irregular and peculiar systems),
with the remainder of sources in hosts with $T < 1$ constituting the
elliptical sample (which includes both elliptical and lenticular host
galaxies).

\begin{figure*}
\begin{center}
\rotatebox{0}{
{\includegraphics[width=475pt]{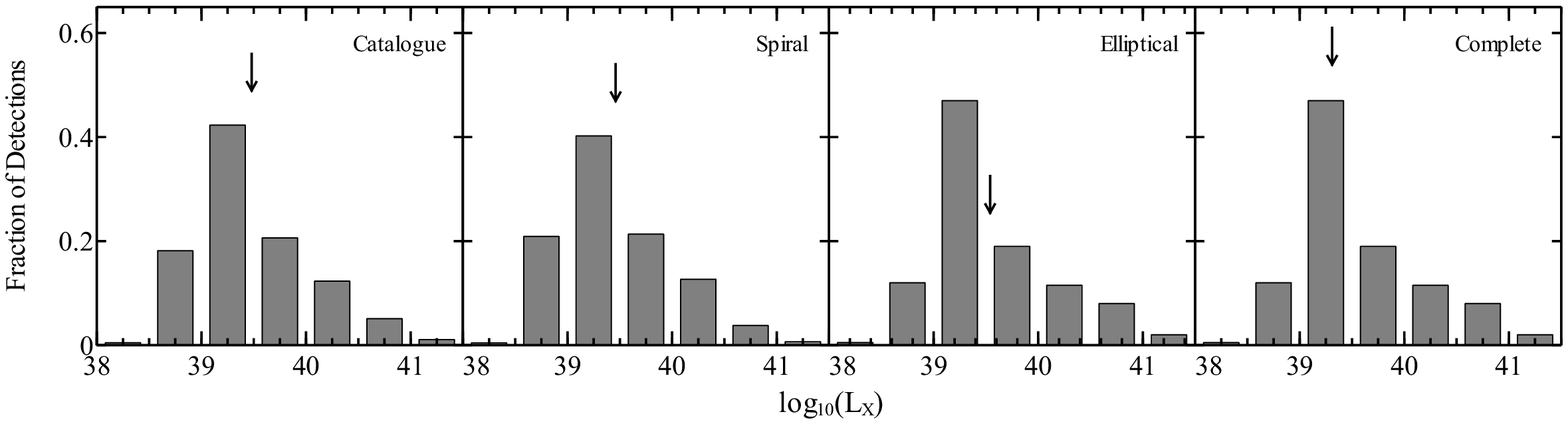}}
}
\rotatebox{0}{
{\includegraphics[width=475pt]{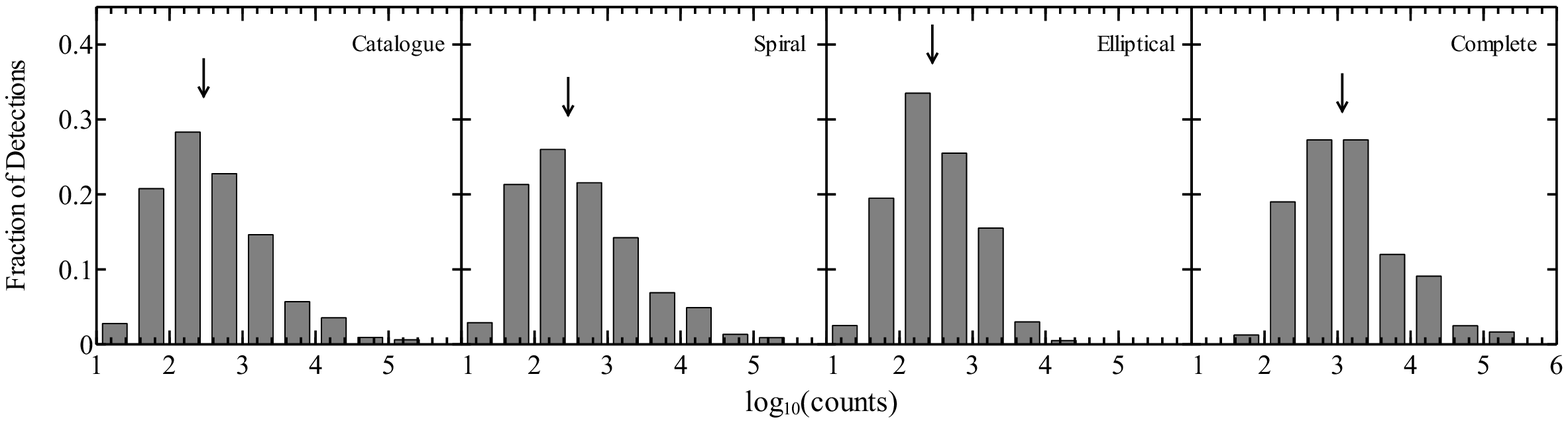}}
}
\end{center}
\caption{Fractional distributions for the X-ray luminosities and the
number of photons detected from the ULX candidates (0.5 dex bins)
included in the presented catalogue. Arrows indicate the values quoted
in Table \ref{catsum}. The mean X-ray luminosity and the median numbers
of counts received are similar for detections associated with spiral
and elliptical galaxies.}
\label{fig_det_hist}
\end{figure*}

Table \ref{catsum} shows that we detect a factor of $\sim$2 more ULX
candidates in spiral galaxies than in ellipticals, and we have a
slightly higher proportion of multi-epoch datasets for spiral
candidate ULXs.  We also have a factor $\sim$2 more host spiral
galaxies than ellipticals, although as shown in Fig. \ref{fig_gal_hist}
the elliptical hosts have a mean distance $\sim 70\%$ greater than
spirals, and also are on average bigger and brighter with a mean
absolute magnitude nearly half a magnitude higher. The mean log
X-ray luminosity of the candidate ULX detections in each sample are
approximately the same (see Fig. \ref{fig_det_hist}); however, given
the greater distance to the ellipticals the candidate ULXs possess a
fainter median X-ray flux. Despite this, we accumulate similar median
numbers of X-ray photons per detection in both samples (also
demonstrated in Fig. \ref{fig_det_hist}), since the elliptical
galaxies typically have longer observations. Note that the values
quoted in Table \ref{catsum} are calculated for all the detections
included in the catalogue, if we limit ourselves to detections above
$10^{39}$\,\ergps, the mean log($L_X$) values increase to $39.61 \pm
0.03$ and $39.63 \pm 0.04$ for spiral and elliptical detections
respectively, which are extremely similar.

\begin{table}
  \caption{Energy ranges covered by the 2XMM energy bands.}
\begin{center}
\begin{tabular}{p{0.1cm} c p{1cm} c p{0.1cm}}
\hline
\\[-0.25cm]
& 2XMM Band & & Energy & \\
\\[-0.3cm]
& & & (\kev) & \\
\\[-0.25cm]
\hline
\\[-0.25cm]
& 1 & & 0.2--0.5 & \\
\\[-0.2cm]
& 2 & & 0.5--1.0 & \\
\\[-0.2cm]
& 3 & & 1.0--2.0 & \\
\\[-0.2cm]
& 4 & & 2.0--4.5 & \\
\\[-0.2cm]
& 5 & & 4.5--12.0 & \\
\\[-0.25cm]
\hline
\end{tabular}
\label{tab_energies}
\end{center}
\end{table}

\begin{figure*}
\begin{center}
\rotatebox{0}{
{\includegraphics[width=475pt]{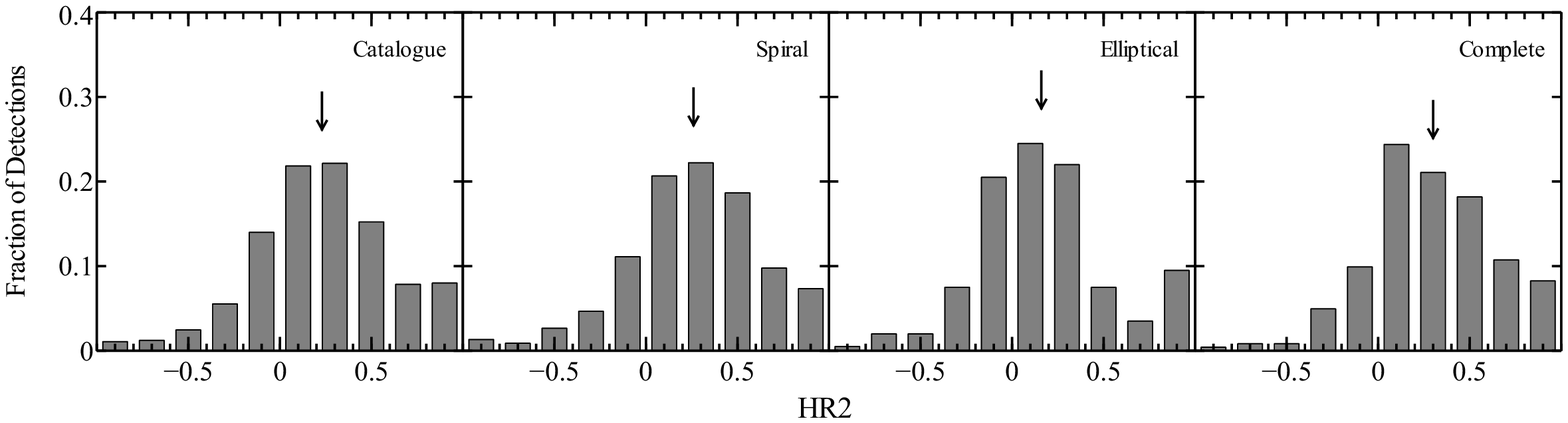}}
}
\rotatebox{0}{
{\includegraphics[width=475pt]{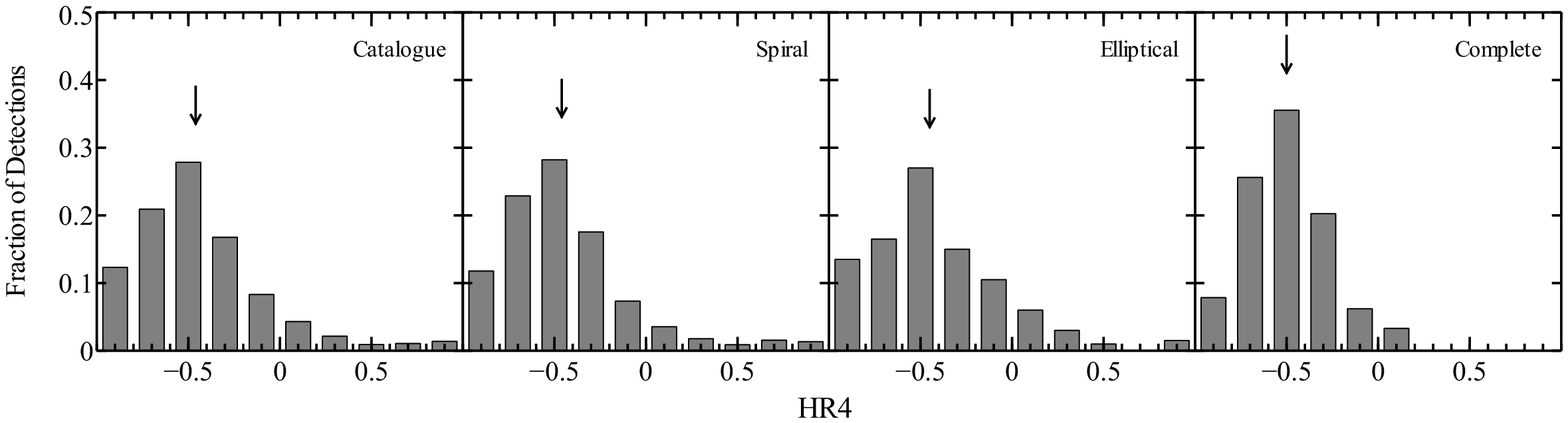}}
}
\end{center}
\caption{Fractional distributions for the 2XMM EPIC hardness ratios HR2 and
HR4 ($\Delta$HR = 0.2 bins) for the ULX candidates included in the catalogue.
Arrows indicate the values quoted in Table \ref{catsum}. Detections from
spiral galaxies appear harder at low energies than their elliptical
counterparts, but this discrepancy is not seen at higher energies. This may
be an indication that the sources in spiral galaxies are more absorbed.}
\label{fig_HR_hist}
\end{figure*}

Perhaps the most interesting distinction between the samples comes in
terms of the 2XMM X-ray hardness ratios, defined as:

\begin{equation}
HR = \frac{C_B - C_A}{C_B + C_A}
\label{eqn_2xmmHR}
\end{equation}

\noindent{where $C_A$ and $C_B$ are the count rates of the two bands under
consideration. The energies covered by the five 2XMM energy bands are given
in Table \ref{tab_energies}; HR1 combines bands 1 and 2, HR2 combines bands 2
and 3, \etc\ The 2XMM hardness ratios are normalised to be in the range [-1,1]
with harder spectra having more positive hardness ratios. Although Fig.
\ref{fig_HR_hist} shows the detections display a wide range of hardness ratios,
the mean hardness
ratio values are constrained to about $\pm 0.02$ in each case; hence the
differences between the spiral and elliptical sample are significant at
the $\sim 3-4.5 \sigma$ level for hardness ratios HR1, HR2 and HR3. In each
case the spiral hardness ratio indicates the sources are somewhat spectrally
harder than their elliptical counterparts, with this spectral distinction
not apparent in HR4. To demonstrate this we show the number distributions
of HR2 and HR4 in Fig. \ref{fig_HR_hist} for both the elliptical and spiral
galaxy detection populations. The difference in the mean values for HR2 is
largely driven by an absence of detections with hard spectra below 2\,\kev\
from elliptical galaxies. We suggest this may primarily be due to the
presence of more neutral gas in spiral galaxies, particularly in the
star-forming regions hosting the ULXs, which will lead to more absorbed and
hence harder spectra. However, a higher column physically associated with
each ULX in spiral galaxies - perhaps a result of mass-loss from the likely
high-mass stellar secondaries in such systems - is also an intriguing
possibility. Whatever the cause, this difference supports the decision to
separate our candidate ULXs into two distinct samples.}

Within the catalogue there are also 5 sources that appear to display
HLX luminosities, none of which have previously been included in ULX
catalogues to date. The brightest of these is 2XMM J134404.1-271410,
which was observed at $L_X = (2.82 \pm 0.24) \times 10^{41}$
\ergps. However the closest HLX included in our catalogue, and hence
potentially the easiest to study, is 2XMM J011942.7+032421 (hereafter
NGC\thinspace470 HLX1), associated with the relatively nearby ($d
\simeq$34 \mpc) spiral galaxy NGC 470.  This source was observed to
reach a luminosity of $L_X = (1.53 \pm 0.08) \times 10^{41}$
\ergps. These sources will be the subject of forthcoming work
(\citealt{Sutton10}; Sutton \etal in prep.), which will include recently
obtained follow-up observations of NGC 470 HLX1.

\subsubsection{The complete sub-sample}
\label{sec_compl}

In order to undertake certain statistical studies of the population,
it is necessary to define a `complete sample', \ie the subset of the
catalogue compiled from observations of galaxies in which all the
ULX candidates present within that galaxy should be detected. By doing
so, we avoid including an artificial bias for brighter sources. Note
that in this context we consider an observation of a galaxy to be
complete if it is possible to detect all the sources that match our
selection criteria; this does not address the limitations with respect
to completeness inherent within these criteria, which are discussed in
\S \ref{sec_lim}.  

To determine whether an observation of a galaxy can be considered complete
there are two limiting flux sensitivities that must be considered and
compared. The first is the sensitivity of the observation, $f_{obs}$, i.e.
the minimum count rate that a source must have to be detected at a certain
position in the detector with a given detection significance. This value is
mainly determined by the local effective exposure and background level. In
order to compute a sensitivity map for each \xmm observation in the
soft (0.5--2.0 keV) and hard (2.0--12.0 keV) bands we used an empirical
approach (see \citealt{Carrera07} and \citealt{Mateos08} for a detailed
description of the method). Briefly, if Poisson statistics hold, it is
possible to determine the minimum count rate that a source must have to be
detected with a given significance. This Poissonian count rate is calculated
from the total number of background counts and mean effective exposure around
the source position. However, source detection and parameterisation in the
2XMM pipeline is a complicated process that involves a simultaneous maximum
likelihood fit of the distribution of source counts detected with each EPIC
camera in different energy bands convolved with the telescope's point spread
function (PSF). Therefore the count rate values resulting from the source
detection process are bound to deviate from the assumption of pure Poissonian
statistics. However, as it is shown in \cite{Carrera07} and \cite{Mateos08},
there exists a linear relationship between Poissonian count rates and those
derived from the \xmm pipeline. This relationship can be used to
empirically correct the pipeline count rates from all non Poissonian effects.
This approach is used by the Flux Limits from Images from \xmm (FLIX)
server provided by the \xmm Survey Science Centre
(SSC)\footnote{http://www.ledas.ac.uk/flix/flix.html}.  Throughout this work,
we utilise sensitivity maps calculated for a 5$\sigma$ detection significance.

\begin{figure}
\centering
\includegraphics[width=8.3cm,angle=0]{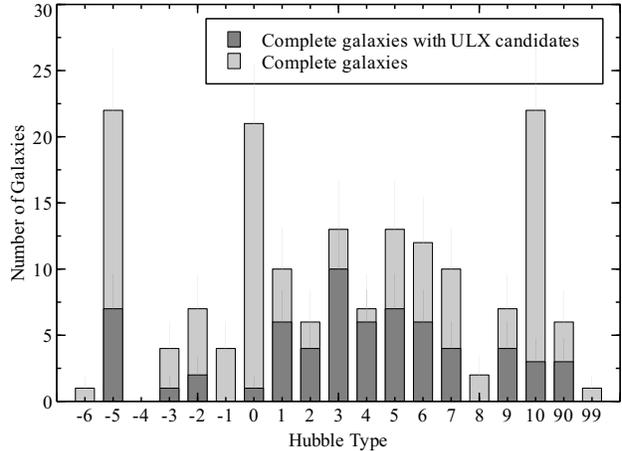}
\caption{Number distributions of the Hubble type for all galaxies
with complete observations, and galaxies with complete observations
that host one or more ULX candidates. It is clear that ULXs are
preferentially located in early type spiral galaxies.}
\label{fig_compl_T}
\end{figure}

The second is the flux of a source emitting at $10^{39}$ \ergps\ at the
distance of the galaxy in question, which we refer to as the ULX luminosity
limited sensitivity, $f_{ulx}$. In order to consistently compare $f_{obs}$
and $f_{ulx}$, we calculate $f_{ulx}$ for the full \xmm bandpass (0.2--12.0\,\kev),
then distribute the sensitivity between the hard and soft bands according to an
average ULX spectral shape. We adopt a simple absorbed power-law model, with
$\langle \Gamma \rangle = 2.2$ and $\langle$\nh$\rangle = 2.4\times10^{21}$
\atpcm, determined from the sample presented in \cite{Gladstone09}. Using
PIMMS\footnote{http://heasarc.nasa.gov/docs/software/tools/pimms.html} we
find that approximately 36 per cent of the flux from ULXs should be observed
in the soft band.

Having calculated $f_{ulx}$ for each energy band, we compare this
with $f_{obs}$ for each observation of each galaxy across its projected
area (excluding chip gaps, \etc, as well as the circular region of radius
7.5'' excluded from our analysis during our consideration of nuclear
sources). This comparison is performed separately
for the \epicpn\ and two \epicmos\ detectors. An observation of a
galaxy in a particular band with a particular detector is considered
complete if $f_{ulx}$ is greater than $f_{obs}$ for the whole of the
considered galaxy area. We take an observation of a galaxy to be
complete in general if it can be considered complete in either of the
bands for any of the detectors, as regardless of whether it is
complete for only one band in one detector or both bands in all
detectors, all the ULXs detectable with our selection criteria should
appear in the catalogue. On investigation, we find that any
observation that is considered complete in the hard band is always
considered complete in the soft band as well.

The complete sub-sample consists of 242 detections of 169 discrete
sources, including 130 sources which are observed to radiate at or
in excess of $10^{39}$ \ergps. Of these, 106 are found in 52 spiral
galaxies, and 24 in 12 elliptical galaxies. The spatial coverage of
2XMM includes complete observations of a total of 108 spiral and 56
elliptical RC3 galaxies (i.e. galaxies for which the completion
criteria are met, regardless of whether they actually host a ULX
candidate). In Fig. \ref{fig_compl_T} we show the number
distributions of the Hubble type for all the galaxies with complete
observations, as well as those with complete observations that host
ULX candidates. It is clear that ULXs are preferentially located in
early type spiral galaxies, as would be expected given the
association between ULXs and regions of high star formation
(\citealt{Swartz09}). The lack of ULX candidates in elliptical
galaxies is a consequence of their relatively low star formation
rates, while the lack of ULX candidates in late type spirals arises
due to the majority of these being dwarfs galaxies, which do not
contain enough stellar mass to host ULXs. Of the 169 sources
included in the complete sample, 106 are `new' detections, of which
71 are observed to radiate at or in excess of $10^{39}$ \ergps.
Finally, there are also 39 sources included that are not observed to
radiate at $L_X \geq 10^{39}$\,\ergps\ at any epoch contributing to
the complete sample, although 8 of these sources are observed at ULX
luminosities at other epochs that contribute only to the general
catalogue. These fainter sources are also primarily observed in spiral
galaxies, with only 7 of the 39 associated with ellipticals.

In Table \ref{catsum} and Figures \ref{fig_det_hist} and
\ref{fig_HR_hist} we provide a comparison of the complete subsample
and the catalogue as a whole. As expected, the mean distance to
host galaxies for the complete sample is significantly smaller than
that of the catalogue as a whole (all galaxies with complete
observations are within 24\,Mpc), and the host galaxies are fainter
on average. In addition, although the detections in the complete
sample typically have many more counts, their average luminosity
is lower than that of the whole catalogue. This arises naturally
from the fact that a large portion of the detections in the full
catalogue arise from observations of galaxies that did not have
the sensitivity to detect sources at 10$^{39}$ \ergps. Consequently,
there is a slight bias towards brighter sources in the full
catalogue, which is not present in the complete subsample. Finally,
there are also some slight differences in the average hardness
ratios between the complete sample and the full catalogue, which
will be discussed in more detail in forthcoming work (Walton \etal
in prep.).

\subsection{Estimating the Contribution from Unknown Contaminants}
\label{sec_bg_contam}

\begin{table*}
  \caption{Fractional background contamination estimates calculated for
the hard and soft bands for the PN and MOS detectors (see text). The
analysis is also broken down into separate considerations of the
catalogue subsets from spiral and elliptical type galaxies, as well as
the complete sub-sample.}
\begin{center}
\begin{tabular}{c c c c c c c}
\hline
\\[-0.26cm]
& \multicolumn{6}{c}{Estimated Contamination (\%)} \\
\\[-0.3cm]
\hline
\\[-0.26cm]
& Soft & Hard & Soft & Hard & Soft & Hard \\
\\[-0.3cm]
\hline
\\[-0.26cm]
& \multicolumn{2}{c}{All Galaxies} & \multicolumn{2}{c}{Spiral Galaxies} & \multicolumn{2}{c}{Elliptical Galaxies} \\
\\[-0.3cm]
\hline
\\[-0.26cm]
PN & $33.9\pm4.6$ & $12.9\pm3.7$ & $28.3\pm4.8$ & $10.8\pm3.5$ & $46\pm10$ & $19\pm10$ \\
\\[-0.3cm]
MOS1 & $35.8\pm5.1$ & $19.5\pm4.2$ & $29.9\pm5.5$ & $12.8\pm4.0$ & $47\pm11$ & $39\pm12$ \\
\\[-0.3cm]
MOS2 & $37.1\pm5.4$ & $18.5\pm4.1$ & $28.8\pm5.3$ & $12.5\pm4.0$ & $58\pm14$ & $35\pm11$ \\
\\[-0.3cm]
\hline
\\[-0.3cm]
Weighted & \multicolumn{2}{c}{\multirow{2}{*}{$24.0\pm4.1$}} & \multicolumn{2}{c}{\multirow{2}{*}{$17.9\pm3.8$}} & \multicolumn{2}{c}{\multirow{2}{*}{$39\pm5$}} \\
Average \\
\\[-0.3cm]
\hline
\\[-0.26cm]
& \multicolumn{2}{c}{All Complete Galaxies} & \multicolumn{2}{c}{Complete Spiral Galaxies} & \multicolumn{2}{c}{Complete Elliptical Galaxies} \\
\\[-0.3cm]
\hline
\\[-0.26cm]
PN & $32.9\pm6.6$ & $14.0\pm4.6$ & $28.0\pm6.5$ & $11.8\pm4.5$ & $54\pm21$ & $23\pm15$ \\
\\[-0.3cm]
MOS1 & $37.2\pm7.3$ & $17.9\pm5.0$ & $31.9\pm7.3$ & $14.1\pm4.9$ & $58\pm22$ & $34\pm17$ \\
\\[-0.3cm]
MOS2 & $35.8\pm7.1$ & $17.5\pm5.0$ & $30.5\pm7.1$ & $13.8\pm5.0$ & $57\pm22$ & $32\pm16$ \\
\\[-0.3cm]
\hline
\\[-0.3cm]
Weighted & \multicolumn{2}{c}{\multirow{2}{*}{$22.5\pm4.4$}} & \multicolumn{2}{c}{\multirow{2}{*}{$18.5\pm4.0$}} & \multicolumn{2}{c}{\multirow{2}{*}{$39\pm6$}} \\
Average \\
\\[-0.3cm]
\hline
\end{tabular}
\label{tab_bg_contam}
\end{center}
\end{table*}

When producing a catalogue of any specific astronomical object it is
important to consider what fraction of the sources included are likely
to be undesirable contamination, \eg background quasars in this
case. Here we attempt to quantify the expected contribution from such
unknown contaminants. In brief, this requires estimation of the
number of background sources that should be observed at some minimum
detection significance (for \textit{all} RC3 galaxies observed), and
comparing this with the actual number of sources detected at or above
that minimum significance. Like the consideration of observation
completeness, our calculations have been performed separately for each
EPIC detector in the soft and hard bands, as dictated by the
availability of the sensitivity maps.

Since the catalogue compiled is of sources with $L_X > 10^{39}$
\ergps, only background contaminants with apparent luminosities also
in this range needed to be considered, so we again make use of the
full band $f_{ulx}$ calculated previously. However, we are now
considering a different class of source, most likely background
quasars, which on average may not have their flux distributed between
the soft and hard bands in the same way as ULXs. \cite{Piconcelli03}
find that observed spectra of the majority of background quasars are
adequately modelled with a powerlaw continuum, modified by Galactic
absorption, so we again adopt a simple absorbed power law model for the
average spectral shape of background sources. In this calculation
we use the average photon index of $\langle \Gamma \rangle = 1.59 \pm 0.02$
obtained by \cite{Piconcelli03} and the Galactic column density in the
direction of each galaxy (\citealt{NH}). For each galaxy the fraction of
the 0.2--12.0\,\kev\ flux that should be observed in the soft
and hard bands (0.5--2.0 and 2.0--12.0\,\kev\ respectively) was determined
using PIMMS, and $f_{ulx}$ was recalculated for each band accordingly. On
average $\sim$30 per cent of the flux from these sources should be observed in
the soft band. We again compare these sensitivities to the observation
sensitivities provided by the sensitivity maps, $f_{obs}$. The true limiting
sensitivity $f_{lim}$ relevant to this calculation is whichever of these two
is greater, as expressed by equation \ref{eqn_flim}:

\begin{equation}
 f_{lim}=\left\{\begin{matrix}
 f_{ulx} & f_{ulx} > f_{obs}\\ 
 f_{obs} & f_{ulx} < f_{obs}
\end{matrix}\right.
\label{eqn_flim}
\end{equation}
\vspace{0.1cm}

\noindent{Adopting this form simultaneously accounts for observations of
galaxies deep enough to detect sources fainter than ULXs, and
observations in which not all the ULXs in a galaxy could be detected.
Using equation \ref{eqn_flim}, modified sensitivity maps of $f_{lim}$ were
generated.}

It has long been known that the majority of the Cosmic X-ray
Background (CXB) can be accounted for by emission from discrete
sources, as summarized in \cite{Hasinger04}.  \cite{Moretti03}
investigated the number of such sources that should be resolved
($N(>$$S)$, per unit sky area) as a function of flux sensitivity, S,
finding that this was well modelled as a smooth broken power law in
both the hard and soft bands (note that the hard band used in
\citealt{Moretti03} only covers 2.0--10.0\,\kev, however we expect
that any corrections due to the slight differences in energy range
between their hard band and the hard band used here will be minimal, so
it should still be acceptable to compare the two). These relations were
used to convert our limiting sensitivity maps into maps of the number
of background contaminants expected from each pixel (one pixel covers
a $4\times4$ arcsec area of the sky). The total background contribution
from each galaxy was taken as the sum of all the individual pixel
contributions that fell within their \d25 isophote (again excluding
the nuclear region, chip gaps, \etc), and the total background
contribution expected in the catalogue was then the sum of
the contributions from all the RC3 galaxies covered by observations
included in 2XMM. Where galaxies have been observed on multiple
occasions, only the contribution from the deepest observation was
considered.

In order to compare the number of expected background sources with the
number included in the catalogue we generated hard and soft band
maximum likelihoods ($ML$s) using those given for the five basic
energy bands included in 2XMM. We make the approximation that the hard
and soft $ML$ values are the sum of the likelihoods of the basic band
they each encompass, which is increasingly good for higher detection
likelihoods. We then counted the number of sources (note that we
counted \textit{sources} rather than detections, as we only considered
the deepest observation for galaxies observed multiple times) with a
detection significance of at least 5$\sigma$ ($ML > 15$). These were
compared with the number of estimated background sources calculated
previously to obtain fractional contamination estimates for each band
and each detector, after taking into account the known background
contaminants already removed (see \S \ref{sec_clean}). Our estimates
are given in Table \ref{tab_bg_contam}, with quoted uncertainties
based on counting statistics. We also present the results of the same
calculation for the complete sub-sample of the catalogue, as defined
in \S \ref{sec_compl}.

The average contamination is found to be $\sim$24 per cent, similar to
that estimated for previous ULX catalogues (\citealt{Swartz04}). It is
clear from Table \ref{tab_bg_contam} that the subset of detections
associated with elliptical galaxies should be significantly more
contaminated by unidentified background sources than the detections
associated with spiral galaxies. Such a result may have been expected
given the known observational characteristics of the ULX population
which, as highlighted in \S \ref{sec_cat}, are also reflected in this
catalogue.  ULXs are often associated with star forming regions
(\citealt{Swartz09}), which are more common to spiral galaxies than
elliptical galaxies, hence ULXs are more commonly observed in spiral
galaxies. In addition, the observed elliptical galaxies are in
general both larger and further away than the observed spiral galaxies,
hence covering a greater cumulative area on the sky at lower ULX-detection
fluxes. This combination leads naturally to a higher estimation of the
fractional contamination.

Finally, we note briefly that the average contamination estimates
presented in Table \ref{tab_bg_contam} may be better considered as
upper limits to the fractional contamination rather than true
predictions of the likely value. This is because our calculation
does not correct for the flux extinction that will be suffered by
background sources due to the gas and dust in the galaxies these
sources are falsely associated with. Such considerations are
especially relevant for spiral galaxies, which should contain higher
columns of neutral gas and dust than their elliptical counterparts,
as suggested by the hardness ratio comparison discussed in section
\ref{sec_cat}. As such, although we quote values based on the
calculations for both the soft and hard bands to be conservative,
the contamination estimates for the hard band, in which effects due
to absorption will be greatly reduced, are likely to be more
representative of the true contamination. This is therefore likely
to be closer to $\sim$17 per cent for the whole catalogue, and as
low as $\sim$12 per cent for the spiral galaxy population.

\subsection{Limitations of the Catalogue}
\label{sec_lim}

The largest dedicated catalogue of ULX candidates compiled to date is
presented in \cite{Swartz04}, and includes 154 discrete sources with
an estimated 25 per cent background contamination. The catalogue
presented here, with 475 sources and at most a similar contamination
represents a significant improvement in number\footnote{Since this
work was submitted, a large catalogue of general X-ray point sources
in nearby galaxies has been published by \cite{JLiu11}, compiled from
\chandra\ observations. That catalogue contains 300 candidate ULXs
within the D$_{25}$ galaxy regions, and an additional 179 candidates
in the region between the D$_{25}$ and 2D$_{25}$ ellipses, although
we caution that a large fraction ($\sim$60 per cent) of the latter
sample are likely to be contaminats.}, and should provide a
useful resource for the ongoing investigations into the properties of
the ULX population; see \S \ref{sec_science} for examples of possible
analyses that may be performed. However, before embarking upon any
such study it is important to consider the limitations of this
catalogue.

The fractional contamination could be further reduced by highlighting
and removing additional contaminants with multi-wavelength
observations and correlations. Such analysis is beyond the scope of
this paper, but could be performed in future work involving follow-up
observations. \cite{Wong08} present an example of highlighting
background AGNs with optical observations, but spectra from other
wavelengths could also be used to detect characteristic features of
AGNs, e.g. broad and narrow emission lines from Seyfert galaxies.

Perhaps the major limitation on this catalogue is its incompleteness.
There will inevitably be a population of ULXs residing in the galaxies
observed that are not included in this catalogue. Obviously the
observations of some galaxies were not deep enough to detect all the
sources down to $10^{39}$ \ergps. In addition, a number of legitimate
ULXs will have been discarded in the empirical nuclear cut and by
discarding extended sources.  In terms of the latter reason, star
forming regions often appear as diffuse, extended emission and due to
their well documented association with these regions it is likely some
ULXs will be embedded within them. With its fairly modest spatial
resolution, \xmm can have difficulty resolving ULXs in this situation,
so by excluding extended sources we are also excluding any ULXs that
may be `hidden' within them. A particularly good example of this
problem for \xmm is the relatively nearby ($d \sim 35$ Mpc), intense
star forming galaxy NGC 3256.  \xmm observations reveal an extended
X-ray emission region centred on the galaxy, with a solitary ULX
outside this region (\citealt{Jenkins04}), whereas \chandra\ has
resolved the central regions to reveal a population of ULXs
embedded within regions of strong diffuse emission (\citealt{Lira02}).
In addition, detections that appear as extended with \xmm may actually
be emission from an unresolved population of point sources. This may
even be the case for a number of detections that appear as point
sources with \textit{XMM-Newton}, a possibility which becomes
increasingly likely for more distant galaxies. The combination of
these effects is difficult to quantify, but the net result is likely
to be that there are additional ULXs in some of the galaxies observed
that are not included in this catalogue, even for galaxies that have
observations we have considered as `complete'. Observations of these
galaxies with \chandra\ should go some way to addressing these issues,
owing to its significant improvement on the angular resolution of
\textit{XMM-Newton}.

\section{Example applications of the catalogue data: analysis and discussion}
\label{sec_science}

In order to demonstrate the scientific potential a resource such as
this catalogue offers, we now present a brief analysis highlighting
some characteristics of the derived population based on an analysis
of the statistically complete sample. We note that the main advantage
enjoyed by the 2XMM catalogue over comparable resources is the higher
photon statistics per detection, enabling more detailed studies of
source characteristics over a broader energy range. This will be
explored in future work (Walton \etal in prep.).

\subsection{Luminosity Function}

One way in which we may investigate the ULX population is through
construction of its luminosity function. In order to avoid the
estimation of corrections to account for incompleteness, we make use
of the previously defined complete subset of the presented catalogue,
for which such corrections should not be necessary, or at least
negligible (even these data may suffer from some level of
incompleteness as we are limited by the \xmm angular resolution; see
\S \ref{sec_lim}). There are a number of galaxies contributing to the
complete sample that are found to have multiple observations that can
be considered complete. Although there is some evidence that X-ray luminosity
functions don't change significantly from epoch to epoch, even if the sources
contributing are variable (see \eg \citealt{Zezas07}), we adopt a
conservative stance and only consider the longest observation of each galaxy.

These observations contribute 121 discrete sources radiating at or
above $10^{39}$ \ergps. We group these into equally spaced 0.1 dex
logarithmic luminosity bins, with the uncertainty on the number of
sources in each bin given by Poisson counting statistics. Using the
SHERPA modelling package\footnote{http://cxc.harvard.edu/sherpa/}, we
fit the luminosity function by minimizing a form of the Cash statistic
($C$; \citealt{cstat}) adapted to be an analogue to \chisq\ (\ie a
reduced statistic of \rcstat $\sim$1 for acceptable fits), and find
that the data are well modelled with a single power-law up to the
highest luminosity sources in the complete sub-sample ($L_X \simeq
5.5\times10^{40}$ \ergps); the differential and integral forms of such
a relation are given in equations \ref{eqn_diff} and \ref{eqn_int}:

\begin{equation}
\frac{dN}{dL_{X}} \propto L_{X}^{-\alpha}
\label{eqn_diff}
\end{equation}

\begin{equation}
\Rightarrow N(>L) \propto L_{X}^{-(\alpha-1)}
\label{eqn_int}
\end{equation}
\vspace{0.075cm}

\begin{figure}
\centering
\includegraphics[width=8.0cm,angle=0]{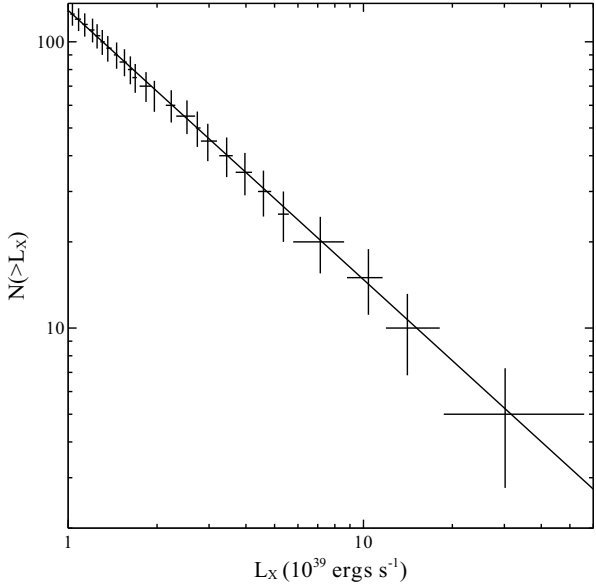}
\caption{The luminosity function derived from the complete subset of the
catalogue. The best fit single power-law model (see text) is also shown.}
\label{fig_lx}
\end{figure}

\noindent{This simple model gives a reduced statistic of \rcstat =
0.89, with the best fit value of the exponent $\alpha$ = $1.96 \pm
0.11$ (here and throughout this section quoted uncertainties are
the 1$\sigma$ uncertainties on one parameter of interest). On
investigation, we find that the quality of fit, and the exponent
obtained are not dependent on our choice of luminosity bin size.
In Fig. \ref{fig_lx} we show the integral form of the luminosity
function (equation \ref{eqn_int}), rebinned so that there are
approximately 5 additional sources from one luminosity bin to the
next for clarity, and the best fit power-law model. The bin
centroids in Fig. \ref{fig_lx} are the average of log($L_{X}$) for
the additional sources included in each bin.}

\begin{figure}
\centering
\includegraphics[width=8.0cm,angle=0]{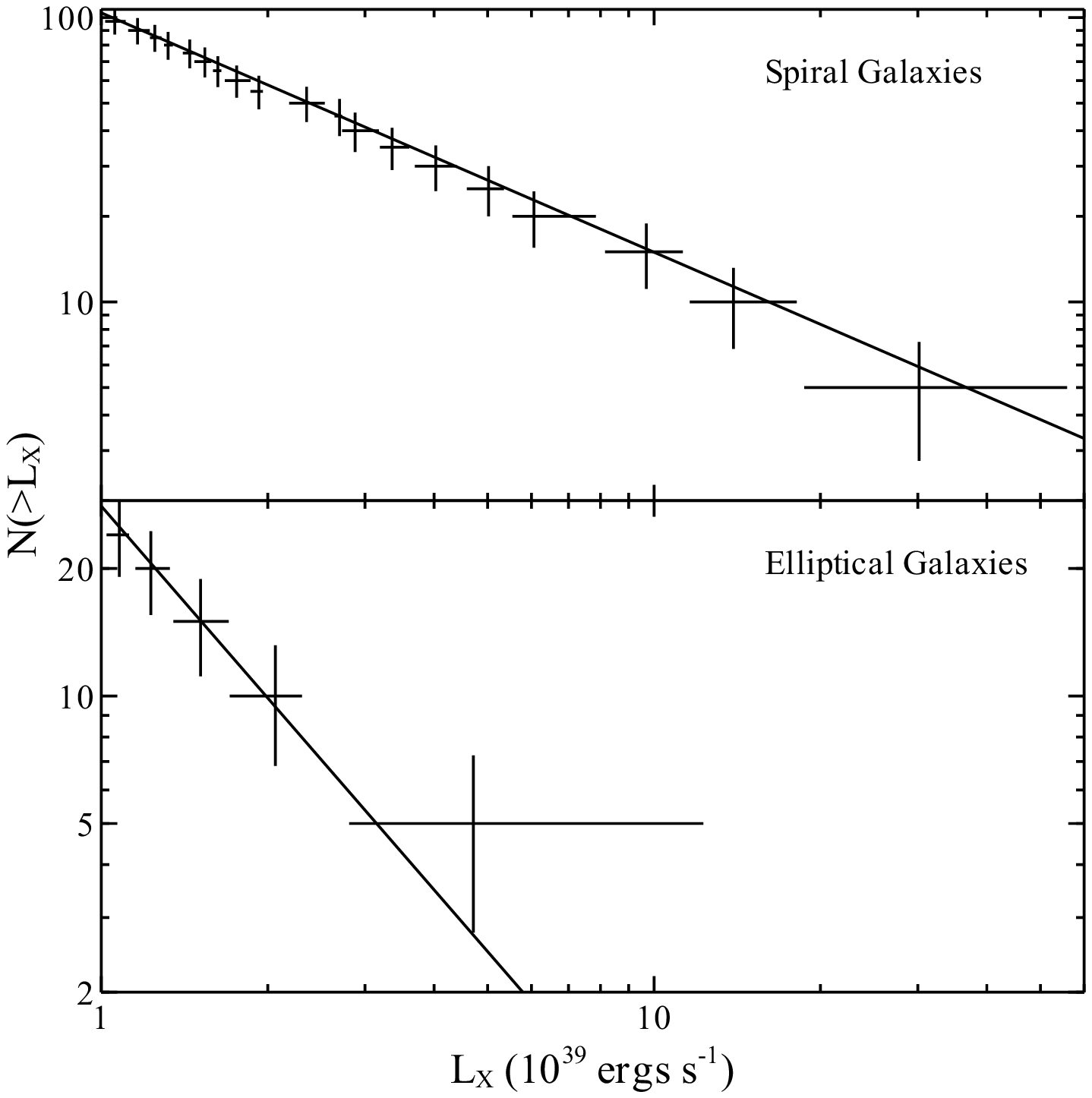}
\caption{The luminosity functions for sources in the complete subset located
in spiral galaxies (top) and elliptical galaxies (bottom). Again, the best
fit power-law model to each is also shown.}
\label{fig_lx2}
\end{figure}

We also investigate the luminosity functions for the spiral and
elliptical galaxy ULX populations separately. Of the 121 sources
considered previously, 97 are located in spiral galaxies and 24 in
elliptical galaxies.  Again, we apply a simple power-law model
to each, and find that this provides an excellent representation of
the data in both cases (\rcstat = 0.97 and 0.7 respectively), with
$\alpha_{spiral}$ = $1.85\pm 0.11$ and $\alpha_{elliptical}$ = $2.5
\pm 0.4$. As with the full complete sample shown in Fig. \ref{fig_lx},
in Fig. \ref{fig_lx2} we show the integral forms of the spiral and
elliptical luminosity functions, rebinned in a similar
fashion. Although the exponents for spiral and elliptical galaxies
agree within their 2$\sigma$ uncertainties, due to the large uncertainty
on the elliptical exponent, which in turn is a consequence of the
relative lack of sources from those galaxies, there is a tentative
suggestion that the luminosity function for sources in elliptical
galaxies is steeper than for those in spiral galaxies. This would be
consistent with the steepening of the elliptical luminosity function
seen in the samples of \cite{Swartz04} and \cite{Colbert04}. Such a
result would not be surprising given the expectation that the sources
in elliptical galaxies are likely to be LMXBs, while those in spiral
galaxies are likely to be HMXBs (\citealt{Colbert04};
\citealt{Humphrey03}), and the shapes of the `universal' luminosity
functions compiled for these two classes of BHB, see \cite{Grimm03}
and \cite{Gilfanov04c}. The exponents obtained here are consistent
within uncertainties with the relevant exponents for the HMXB and LMXB
luminosity functions obtained in those works, both of which extend to
luminosities below $10^{39}$ \ergps. This might suggest that the same
process, or combination of processes, responsible for forming the lower
luminosity BHB population are also responsible for forming the majority
of ULXs.

It has previously been reported that there may be a high luminosity
cut-off in the ULX luminosity function at $L_{X}\sim 2\times10^{40}$
\ergps, see \eg \cite{Swartz04} but especially \cite{Grimm03} and
\cite{Gilfanov04b}, who argue strongly for an upper limit to the
luminosity of HMXBs. Given the large overlap between the ULX
and HMXB populations this is also relevant for the ULX luminosity
function. Such a cut-off may arise due to there being an upper limit
to the mass of the black holes HMXBs are expected to contain, which
\cite{Grimm03} estimate to be $\sim$100\msun\ under the assumption
that the Eddington limit is not violated. Although the data are
already reproduced well with a single power-law function, we also
investigate a broken power-law model, as expressed in equation
\ref{eqn_bpl}:

\begin{equation}
\frac{dN}{dL_{X}} \propto \left\{\begin{matrix}
L_{X}^{-\alpha_{1}} & L_{X} < L_{c} \\ 
\\[-0.3cm]
L_{X}^{-\alpha_{2}} & L_{X} > L_{c}
\end{matrix} \right.
\label{eqn_bpl}
\end{equation}
\vspace{0.08cm}

\noindent{However, this more complex model provides a negligible
improvement in the fit to the data: we find $\Delta C$ = 0.1 for 2 extra
degrees of freedom, the break luminosity $L_{c}$ is not well
constrained (although its best fit value is $1.1\times10^{40}$\ergps,
similar to previously proposed values for $L_c$), and the two
exponents are consistent within their 1$\sigma$ uncertainties. Since
the work of \cite{Grimm03} focuses on HMXBs and \cite{Swartz04} only
see the cut-off in the luminosity function of spiral galaxies, we
repeat the comparison using only the sources from complete spiral
galaxies. Unsurprisingly, as our overall luminosity function is
dominated by sources from spiral galaxies, we find exactly the same
results, with an improvement now of $\Delta C$ = 0.2 for 2 extra
degrees of freedom.}

We must remind and caution the reader that our luminosity functions
are generated from sources from ensembles of galaxies. Although the
broad, general shape should remain robust, non-systematic errors in
the estimated luminosities that differ from galaxy to galaxy due to
\eg poorly estimated galaxy distances that could smear out complex
features such as a cut-off. If present, the effect of such
uncertainties would be especially prevalent at high luminosities where
there are fewest sources. In this work we have adopted a simple Hubble
Flow distance for galaxies with $cz > 1000$\kms\ and the distance
quoted in \cite{TULLY} for galaxies with $cz \leq 1000$\kms; a more
careful analysis of galaxy distances, which is beyond the scope of
this work, might clarify this issue. Other biases may also be present
too, for example background contamination. Removing these sources
might alter the shape of the pure ULX luminosity function. However,
a repeat of the work in \S \ref{sec_bg_contam} for sources with $L_X$
above $5 \times 10^{40} \ergcms$ (limited to within 100 Mpc) shows
$\sim$30 per cent of the sources at these luminosities in the catalogue
are likely background contamination, and only $\sim$15 per cent of the
spiral galaxy population (Sutton et al., in prep.). This is
fairly similar to that found for the whole ULX luminosity range, so it
seems that the fractional contamination remains roughly constant with
luminosity, hence removing such sources would primarily effect the
normalisation rather than the shape of the luminosity function.

The luminosity functions presented are still ultimately limited by low number statistics at
the highest luminosities, which severely limits the comparison of
complex models. So, although we conclude that our data do not require
anything more than simple power-law models, we cannot rule out the
presence of an intrinsic cut-off in the ULX luminosity function at
$L_c \sim 2\times10^{40}$ \ergps. However, the complete sub-sample
has 5 sources in its highest bin, which nominally lies above this
break. If the value adopted for $H_0$ is significantly underestimated
then source luminosities might be systematically overestimated, but
this seems unlikely given the currently favoured value of $H_0 =
72 \pm 8$ \H0 (\citealt{H0}). Given the expected level of contamination
it is also unlikely these are all foreground/background sources. They must therefore be
regarded as \textit{bona fide} ULX candidates beyond the previously reported
limit, which must call into question the veracity of this proposed
feature. Indeed, the full catalogue contains an additional 38 sources
above this proposed upper limit, although we note again that \chandra\
follow-up of all of these high luminosity sources is strictly required
to confirm their nature as single point sources.

Even though these additional sources may not be included in our
quantitative analysis of the luminosity function due to
incompleteness, their simple presence in the catalogue carries
interesting implications with respect to the possible presence
of this luminosity cut-off. We first consider the case that
these sources represent a smooth continuation of the universal
HMXB luminosity function observed below $\sim10^{40}$ \ergps.
This luminosity is the Eddington luminosity ($L_E$) for a
100\,\msun\ black hole, or 10\,$L_E$ for a 10\,\msun\ black
hole. One way in which higher luminosities may be explained is
to invoke larger black holes. If there is no upper limit to the
HMXB X-ray luminosity, then their luminosity function no longer
provides evidence for an upper limit to the HMXB black hole mass. 
The situation is complicated by the possibility that some XRBs
appear to radiate at super-Eddington rates (\eg GRS\,1915+105;
\citealt{Done04}), so an upper mass limit could still be present
without manifesting itself clearly in the luminosity function.
\cite{Zampieri09} independently argue that even in low metallicity
environments it is not possible to produce black holes
significantly more massive than $\simeq90$\,\msun\ through
standard stellar evolution. However, a smooth luminosity function
would nevertheless remove an important observational result
supporting such an upper mass limit.

The lack of a high luminosity cut-off would also raise interesting
questions on the nature of the relationship between the total
X-ray luminosity from HMXBs in a galaxy and its star formation
rate presented by \cite{Gilfanov04a} and \cite{Gilfanov04b}.
They argue that an upper limit to the HMXB luminosity function
is required to produce the linear regime of their relation at
high star formation rates, based on the non-trivial statistics
of number distributions that follow a power-law. This linear
regime requires that the probability distribution of the total
HMXB X-ray luminosity is Gaussian, which may only occur for
galaxies with high enough star formation rates such that there
are a fair number of sources at or near the upper limit. These
sources dominate the total HMXB X-ray luminosity and remove the
discrepancy between the mean and the mode values for this
quantity which is otherwise present for populations drawn from
luminosity functions with exponents $1<\alpha<2$, and causes the
non-linear regime at low star formation rates. Without an upper
luminosity limit, this non-linear regime should continue to
higher star formation rates. However, the data do not appear to
continue this trend (\citealt{Gilfanov04b}), which would be an
important discrepancy to address. We note briefly that the data
presented in those works do not account for intrinsic absorption
and obscuration, which may play an important role. Unfortunately,
attempting to account for this invariably leads to model based
assumptions and uncertainties.

We now also consider the case that there \textit{is} a cut-off in
the HMXB luminosity function at $\sim2\times10^{40}$\,\ergps.
As argued previously, although it may be that the luminosities
of some of the sources included in the catalogue that appear
in excess of this limit have been poorly estimated, and some
sources may turn out to be background quasars, this is
unlikely to be the case for all 43. We would then be observing
a population of sources radiating above the maximum HMXB
luminosity. These must represent a different class of black
hole, with a different formation mechanism. There is speculation
that galaxies in the highest star formation regime may produce a
population of IMBHs through black hole mergers in dense stellar
clusters (see \eg \citealt{Gilfanov04b, PortZwart04}), which
could be seen at luminosities in excess of $10^{40}$ \ergps.
Fig. 10 in that work shows the effect such a population might
have on the universal HMXB luminosity function; the cut-off
instead becomes a step-like feature due to the upper HMXB mass
limit, and above this step a far less numerous population of
IMBHs are observed. Given the links between ULXs and star
formation, it might not be unreasonable to argue that by
focusing on galaxies that host ULXs, we are selecting galaxies
with high star formation rates. The presence of such high
luminosity sources in our catalogue therefore might not be
surprising even in the presence of an upper limit to the HMXB
black hole mass, and may be evidence for a rare population of
IMBHs. These intriguing sources are investigated further in
Sutton \etal (in prep.).

\subsection{Specific ULX Frequency}

Here we again make use of the complete sub-sample to mimic the analysis
presented by \cite{Swartz08}, and investigate how the specific ULX
frequency $S^u$, \ie the number of ULXs per unit galaxy mass, evolves as
a function of galaxy mass. This provides additional information on the
type of galaxy, and hence the environments in which ULXs are
preferentially formed. In order to estimate galaxy mass for the galaxies
with complete observations, we make use of the relations between optical
colour and stellar mass-to-light ratio, $M/L$, presented in
\cite{Bell03}, which should hold for both elliptical and spiral type galaxies
(we are assuming that stellar mass is directly proportional to galaxy mass).
The derivation of these relations utilises SDSS data (\citealt{SDSS})
and the P\'{E}GASE galaxy evolution models (updated from those originally
presented in \citealt{Fioc97}), and assumes a scaled version of the
Salpeter Initial Mass Function (the same as adopted in earlier work
by \citealt{Bell01}). We choose to calculate $M/L$ in the $B$-band,
and utilise the $B$-band magnitude and the $(B-V)$ colours provided
in RC3 to estimate the galaxy mass. The scatter around the relation
between $M/L_B$ and $(B-V)$ is estimated to be $\sim$0.1 dex, so we
include this in the estimated uncertainty of our $M/L_B$ values. The
colours and magnitudes used have been corrected for Galactic and
internal extinction, and for redshift, and when converting absolute
magnitudes to solar $B$-band luminosities, we adopt an absolute solar
$B$-band magnitude of 5.47 (\citealt{Cox2000}).

Through this method, stellar mass estimates were obtained for 86 and
35 of the 108 and 56 complete spiral and elliptical galaxies
respectively. Galaxies without mass estimates, which were mainly a
mixture of elliptical and small nearby galaxies (in particular dwarf
spheroidal galaxies), were discarded from
our analysis. For each galaxy with a mass estimate, we counted the
number of ULX candidates assumed to be associated with it. As with the
luminosity function analysis, we attempt to minimize the effects of
transient and highly variable sources by considering only sources
detected in the longest observation of each galaxy. Then, grouping all
the complete galaxies into equally spaced decade mass bins, we
calculate $S^u$ for each bin by dividing the total number of ULX
candidates observed by the total galaxy mass contained within that
bin. This is expressed in equation \ref{eqn_ulxfreq} for the
$i$th bin containing $n_i$ galaxies:

\begin{equation}
S_{i}^{u}=\frac{N_i}{M_i}=\frac{\sum_{j=1}^{n_i}N_j}{\sum_{j=1}^{n_i}M_j}
\label{eqn_ulxfreq}
\end{equation}
\vspace{0.08cm}

\begin{figure}
\centering
\includegraphics[width=7.6cm,angle=0]{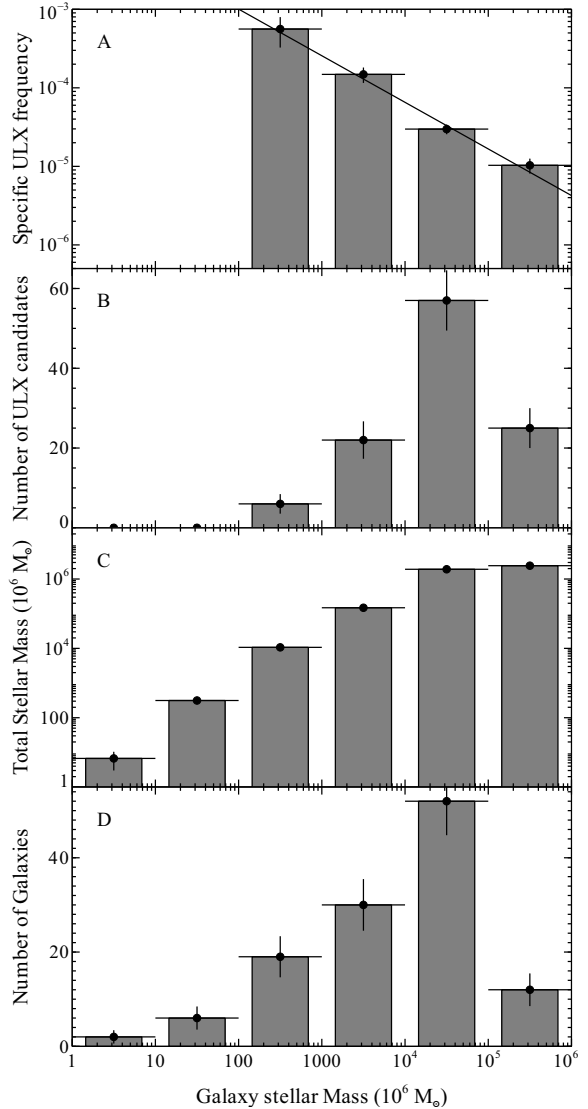}
\caption{The specific ULX frequency and the best fit power-law model
(panel A; see text), total observed
ULX candidate detections (panel B), total stellar mass and number
of galaxies (panels C and D) as a function of galaxy stellar mass for
galaxies with complete observations included in 2XMM.}
\label{fig_ulxfreq}
\end{figure}

\noindent{where $N_i$ and $M_i$ are the total number of ULXs and the
total galaxy mass in bin $i$ respectively, $N_j$ is the number of
ULXs associated with the $j$th galaxy within that bin, and $M_j$ is
the mass of that galaxy. Fig. \ref{fig_ulxfreq} shows the specific
ULX frequency plotted as a function of galaxy mass, as well as the
number of galaxies contributing to each bin, the number of ULX
candidates in each bin, and the total galaxy mass in each bin.}

\begin{figure*}
\centering
\includegraphics[width=15.5cm,angle=0]{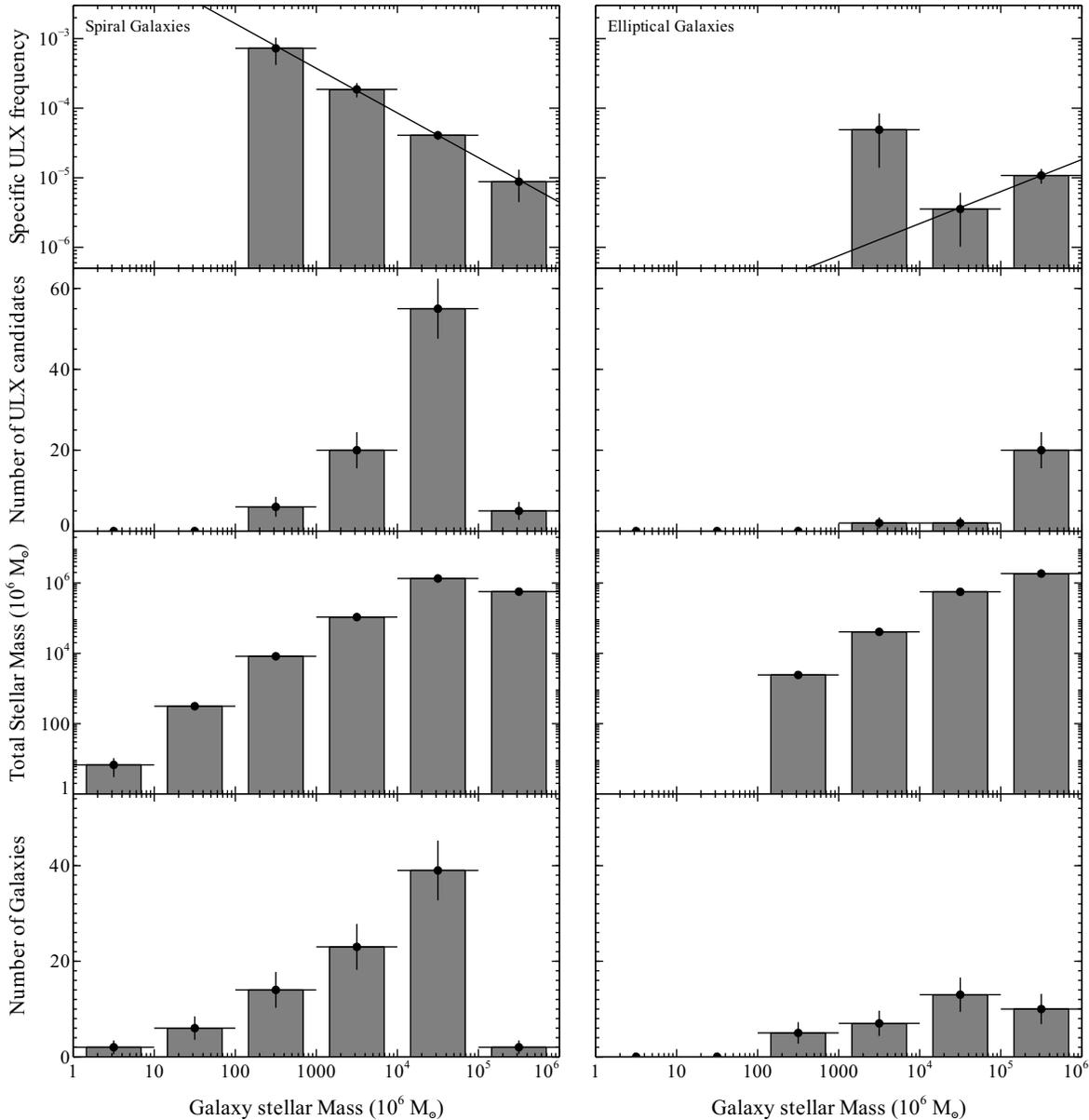}
\caption{The specific ULX frequency and the best fit power-law model calculated
for spiral (left column) and elliptical galaxies (right column) separately. The
vertical panels are as in Fig. \ref{fig_ulxfreq}.}
\label{fig_ulxfreq2}
\end{figure*}

It is clear from Fig. \ref{fig_ulxfreq} that $S^u$ decreases with
increasing galaxy mass, qualitatively similar to the trend
presented in \cite{Swartz08} with a sample compiled from \chandra,
\xmm\ and \ro\ observations of catalogued galaxies within $\sim$8\,Mpc 
(\citealt{Karachentsev04}). Phenomenologically we fit the data (only
for bins in which ULX candidates have been detected) with a power-law,
\ie $S^u \propto M^{-\beta}$ and obtain an index of $\beta = 0.59 \pm
0.07$ (1$\sigma$ uncertainties). With this relation, it becomes
immediately apparent why no ULX candidates are detected in the few
lowest mass bins. The prediction, by extrapolation, for $S^u$ in the
highest mass bin without a detection is $7\times10^{-4}$ ULXs per
$10^6$ solar masses which, considering the total stellar mass contained
within that bin, equates to less than one ($\sim$0.4) expected ULX
detection. For even lower mass bins, the estimated number of ULX
detections continues to decrease; there is simply not enough galaxy mass
contained in the low mass bins to expect ULXs to be detected. The
lack of detections in these bins is therefore not inconsistent with the
trend observed for the bins \textit{with} ULX detections. We also
calculate $S^{u}$ for spiral and elliptical galaxies separately, as shown
in Fig. \ref{fig_ulxfreq2}, and also fit these data with power-law models.
The indices obtained are $\beta_{spiral} = 0.64 \pm 0.07$ and
$\beta_{elliptical} = -0.5^{+0.3}_{-0.5}$. As is clear from Fig.
\ref{fig_ulxfreq2}, such a model does not provide as excellent a
representation of the data for elliptical galaxies as it does for spiral
galaxies. However, we stress that due to a lack of ULX detections and
galaxies with mass estimates, as well as the higher fractional contamination,
the quality of the elliptical data is rather poor. This is reflected in
the large uncertainty on the elliptical index, which is in fact consistent
with zero, \ie the case where $S^u$ is constant with galaxy mass.
Unsurprisingly, since the complete sample is dominated by sources
associated with spiral galaxies, the results obtained for this population
are very similar to the results obtained for the whole sub-sample.

Before discussing the potential physical origins of these results, it is
important to consider possible biases that may be present.  By
considering only sources detected in the longest observation of each
galaxy, we hope to have minimised the influence of transient sources as
best we can.  Background contaminants may have an important effect here,
but as argued in \cite{Swartz08}, they are more likely to be prominent
in the larger, higher mass galaxies, primarily because they cover a much
greater sky area. This serves to artificially increase the number of ULX
candidates and hence $S^u$ in the higher mass bins, and to flatten the
observed trend, \ie decrease $\beta$. If these sources are important,
the values of $\beta$ presented here may only be lower limits. Other
biases may arise due a combination of our ULX selection criteria and
the fairly modest resolution of \textit{XMM-Newton}. By excluding the
nuclear regions of galaxies and extended sources there will be ULXs in
the observed galaxies that are not included in our catalogue. Since we
have defined the nuclear region with approximately a constant angular
size, a greater physical area will be excluded for more distant
galaxies. This is important as the radial distribution of ULX
candidates peaks toward the centres of galaxies (\citealt{Swartz04};
\citealt{Liu06}). On average, the higher mass galaxies are further
away, so this too will have a more important effect on the higher mass
bins, and the number of ULXs in those bins may be slightly
underpredicted. In terms of the exclusion of extended sources, of
which star forming regions may contribute a large number, both galaxy
star formation rate and stellar mass will determine the number of ULXs
missed through being embedded within them. It is not immediately clear
whether this should have any preferential effect on any region of the
mass scale. Finally, there is also the possibility that some of the
2XMM detections are actually a number of unresolved point sources, but
again it is not clear whether this should have a greater effect on any
range of the mass scale than others. However, while these considerations
may have an effect on the quantitative forms of the relations presented,
given the similar trend presented by \cite{Swartz08} to that found
here for spiral galaxies, we can be confident in the qualitative form
of this result.

It is clear that the number of ULXs per unit stellar mass decreases
with increasing galaxy mass for spiral galaxies. There are two main
physical differences between low mass and high mass spirals that are
likely to be combining to give this effect. The first is that the
specific star formation rate, \ie the star formation rate per unit
mass, is higher on average for lower mass spirals (see Fig. 24 in
\citealt{Brinchmann04}). ULXs are well correlated with star forming
regions (\citealt{Swartz09}), and there are typically a larger number
of ULXs in galaxies with high star formation rates (\citealt{Grimm03}).
The second is that lower mass galaxies typically have lower metallicities
(\citealt{Lee06}). ULXs are often observed to be located in low metallicity
regions (see \eg \citealt{Soria05}, \citealt{Zampieri09}, \citealt{Mapelli09},
\citealt{Mapelli10}), and there are a number of theoretical considerations
that suggest the $M_{BH}>10$\msun\ black holes some ULXs are speculated to
host are easier to form in these regions, both due to the formation of
larger progenitor stars, and reduced mass loss rates from stellar winds
(\citealt{Madau01}; \citealt{Bromm04}; \citealt{Vink01}; \citealt{Heger03}).
This combination of higher star formation per unit mass and lower metallicity
provides a natural explanation for the higher specific ULX frequency in low mass
spirals. However, we stress that this does not necessarily mean ULXs are
commonly \textit{observed} in low mass galaxies, for the same reason that
there are no ULX detections in the very lowest mass bins. Although we find
there are more ULXs per unit mass in lower mass spiral galaxies, a
large number of these galaxies often need to be observed to detect ULXs
by virtue of the fact they do not contain much mass.

Although the elliptical data is consistent with $S^u$ remaining
constant with galaxy mass, we briefly comment on the possibility that
$S^u$ might increase with increasing elliptical galaxy mass as formally
suggested by the exponent obtained, as such a scenario would be very
interesting. When stellar binary systems undergo supernova, they
experience a velocity kick (\citealt{Harrison93}; \citealt{Frail94};
\citealt{Lyne94}), and over the course of the lifetime of a galaxy some
fraction of its binary systems may be ejected. It might be that a larger
fraction of binaries are ejected from lower mass galaxies than higher
mass galaxies owing to their weaker gravitational potentials. This
effect might lead to such a positive trend in elliptical galaxies where
ULXs are expected to be LMXBs, hence star formation and metallicity
effects are likely to be minimal. However, we stress again that the
data are consistent with $S^u$ being constant with stellar mass, which
would be a simpler scenario to explain given that LMXBs are expected to
trace stellar mass (\citealt{Gilfanov04c}).

\section{Conclusions}

Through cross correlation with the RC3 galaxy catalogue, we have
mined the 2XMM Serendipitous Survey to produce a catalogue of ULX
candidates. Our catalogue contains 650 detections of 470 discrete
sources, and we conservatively estimate that at most $\sim$24 per
cent, but more likely $\sim$17 per cent of these sources are
undesirable contaminants, most likely background quasars in this
case. Our catalogue offers a significant improvement in the number
of ULX candidates on that of \cite{Swartz04}, which was previously
the largest uniformly selected, dedicated ULX catalogue published
to date. Obviously one of the key
potential uses of this resource is to unearth and provide basic
information on sources for which more detailed observation and/or
analysis would be both interesting and insightful. One such source
that we draw attention to is NGC 470 HLX1, a detailed analysis of
which will be included in forthcoming work by Sutton \etal (in prep).
In addition to highlighting interesting sources, the catalogue may
also be used to directly analyse what we hope to be a representative
sample of the whole ULX population.

To undertake statistical studies of the ULX population, we
define a `complete' sub-sample of the population taken only from
galaxies with observations sensitive enough that we would expect all
the ULX candidates in those galaxies that meet our selection criteria
to be detected. The limitations of these criteria have been discussed
in detail in \S \ref{sec_lim} and are not repeated here. The luminosity
function of this sample is well modelled with a single, unbroken
power-law of form $N(>$$L_X) \propto L_X^{-0.96 \pm 0.11}$. We do not
find any evidence for a break or cut-off in the luminosity function at
$L_X \sim 2\times10^{40}$ \ergps, as has previously been reported in
\cite{Swartz04} and \cite{Grimm03} and, in the latter case, interpreted
as possible evidence for an upper limit of $\sim$100 \msun\ to HMXB
black hole masses. However, given the possible biases introduced by \eg
non-systematic uncertainties in galaxy distances, we certainly cannot
rule out the intrinsic presence of any such cut-off. Additional \chandra\
observations with increased sensitivity of a greater number of nearby
galaxies would help to address this issue. Similar to results obtained in
previous work, \eg \cite{Swartz04}, there is also tentative evidence
that the luminosity function of ULXs in elliptical galaxies is steeper
than that of ULXs in spiral galaxies, which would be consistent with
the assumption that elliptical galaxy ULXs are primarily low mass
binary systems, while those in spiral galaxies are high mass systems.

We find that the specific ULX frequency $S^u$, \ie the number
of ULXs per unit galaxy mass, decreases with increasing galaxy mass
for ULXs associated with spiral galaxies, and is well modelled with a
power-law of form $S^u \propto M^{-0.64 \pm 0.07}$.  Although the
quantitative form of this relation may be affected by biases such as
background contamination, we are confident that the qualitative trend
is robust, as it is similar to the result presented in \cite{Swartz08}.
This trend may be due to a combination of the increase in specific star
formation rate and the decrease in metallicity with decreasing spiral
galaxy mass (\citealt{Brinchmann04}; \citealt{Lee06}). The specific
ULX frequency for elliptical galaxies is found to be consistent with being
constant with galaxy mass, as expected given that LMXBs should trace stellar
mass rather than star formation (\citealt{Gilfanov04c}).

The population analysis presented here is by no means comprehensive,
rather a number of examples of the work that is possible with the
offered resource. We hope that the construction of this catalogue will
encourage future authors to continue this work and so we make the
catalogue available online with this publication\footnote{Catalogue to be made available online via MNRAS;
available in the meantime via private communication}; other interesting
possibilities may be to re-visit the connection between ULXs and star
formation, and to search for further similarities and/or differences
between ULXs in spiral and elliptical galaxies to confirm their
HMXB/LMXB nature. The successes and limitations of our analysis
highlight the importance of utilising both the \xmm and \chandra\
observatories to discover and study ULXs. A number of the limitations
are due to the more modest angular resolution of \textit{XMM-Newton}.
The improvement in this area offered by \chandra\ is essential to
confirm the point-like nature of ULX candidates, and perhaps resolve
additional sources that \xmm could not. However, the larger effective
area of \textit{XMM-Newton} provides detections with better photon
statistics, particularly at high energies, enabling analysis that
would not be possible with the \chandra\ satellite, \eg detecting
high energy spectral curvature (\citealt{Stobbart06, Gladstone09,
Walton4517, Middleton11}; Walton \etal in prep.).

This work also highlights the importance of serendipity
within astronomy. Catalogues such as 2XMM are vital resources for
unearthing new examples and studying populations of various
astrophysical objects. Based on the success of our compilation, we
anticipate that future releases of \xmm source catalogues, will contain
many more ULXs again, providing a further significant contribution to
improving our understanding of the characteristics and diversity of
the ULX population.

\section*{ACKNOWLEDGEMENTS}

The authors would like to thank the referee for their useful suggestions,
which have helped improve this paper, and also acknowledge the
financial support provided by the UK STFC research council. Some of the
figures included in this work have been produced with the
Veusz\footnote{http://home.gna.org/veusz/} plotting package, written by
Jeremy Sanders. This work is based on \xmm observations, an ESA mission
with instruments and contributions directly funded by ESA member states
and the USA (NASA). In addition, this research has made use of the
NASA/IPAC Extragalactic Database (NED), operated by the Jet Propulsion
Laboratory, California Institute of Technology, as well as the Digitised
Sky Survey (DSS), produced at the Space Telescope Science Institute under
U.S. Government grant NAG W-2166.

\bibliographystyle{mnras}
\bibliography{/home/dwalton/papers/references}

\newpage

\onecolumn

\appendix

\begin{table}\section{Example Catalogue Entries}
\begin{center}
\caption{Some basic information for example entries in the presented catalogue, including the
PGC galaxy identifier, the adopted galaxy distance, the 2XMM source ID, the detection luminosity,
potential matches in previous catalogues (\citealt{Colbert02}: CP02; \citealt{LiuBregman05}:
LB05; \citealt{Swartz04}: SW04; \citealt{LiuMirabel05}: LM05) and flags detailing whether the source
is a `new' ULX candidate (\ie not present in any of the above catalogues) and whether the source
detection is included in the complete subset.}
\begin{tabular}{c c c c c c c c c c}
\hline
\hline
\\[-0.3cm]
PGC & Distance & 2XMM &  Detection & CP02 & LB05 & SW04 & LM05 & New? & Complete \\
& (Mpc) & Source ID & Luminosity & Name & Name & RecNo & Name & & Subset? \\
& & & ($10^{39}$\,erg s$^{-1}$) & & & & & & \\
\\[-0.3cm]
\hline
\\[-0.3cm]
1370 & 45.77 & 2609 & $2.25 \pm 0.77$ & & & & & yes & no \\
1388 & 45.0 & 2677 & $3.27 \pm 0.80$ & & & & & yes & no \\
2248 & 120.96 & 4975 & $5.12 \pm 2.18$ & & & & & yes & no \\
2388 & 58.21 & 5291 & $2.50 \pm 0.76$ & & & & & yes & no \\
2789 & 3.0 & 7278 & $1.35 \pm 0.01$ & & & & & yes & yes \\
2789 & 3.0 & 7278 & $1.18 \pm 0.02$ & & & & & yes & yes \\
2789 & 3.0 & 7278 & $0.38 \pm 0.01$ & & & & & yes & yes \\
2789 & 3.0 & 7278 & $0.87 \pm 0.05$ & & & & & yes & yes \\
2789 & 3.0 & 7337 & $1.62 \pm 0.02$ & & & & NGC\,253 ULX2 & no & yes \\
2789 & 3.0 & 7337 & $1.63 \pm 0.02$ & & & & NGC\,253 ULX2 & no & yes \\
2789 & 3.0 & 7337 & $2.26 \pm 0.04$ & & & & NGC\,253 ULX2 & no & yes \\
2789 & 3.0 & 7337 & $2.47 \pm 0.06$ & & & & NGC\,253 ULX2 & no & yes \\
3453 & 75.29 & 9158 & $14.38 \pm 2.65$ & & & & & yes & no \\
4777 & 34.12 & 12122 & $153.03 \pm 7.98$ & & & & & yes & no \\
4801 & 30.87 & 12156 & $1.83 \pm 1.09$ & & & & & yes & no \\
5098 & 65.65 & 12557 & $14.16 \pm 4.07$ & & & & & yes & no \\
5193 & 27.45 & 12818 & $1.42 \pm 0.70$ & & & & & yes & no \\
5193 & 27.45 & 12821 & $0.99 \pm 0.63$ & & & & & yes & no \\
5283 & 73.84 & 13038 & $10.74 \pm 2.40$ & & & & & yes & no \\
5283 & 73.84 & 13038 & $17.58 \pm 9.63$ & & & & & yes & no \\
5974 & 9.70 & 14174 & $2.73 \pm 0.15$ & & & & NGC\,628 ULX2 & no & yes \\
5974 & 9.70 & 14223 & $1.91 \pm 0.17$ & & & & NGC\,628 ULX1 & no & yes \\
5974 & 9.70 & 14223 & $1.04 \pm 0.15$ & & & & NGC\,628 ULX1 & no & yes \\
6983 & 22.88 & 15962 & $1.57 \pm 0.32$ & & & 5 & NGC\,720 ULX7/8 & no & no \\
6983 & 22.88 & 15962 & $1.69 \pm 0.77$ & & & 5 & NGC\,720 ULX7/8 & no & no \\
7252 & 74.15 & 16228 & $11.10 \pm 3.64$ & & & & & yes & no \\
7584 & 66.47 & 16743 & $13.24 \pm 6.90$ & & & & & yes & no \\
8726 & 71.21 & 18486 & $1.78 \pm 1.59$ & & & & & yes & no \\
8974 & 31.25 & 20178 & $17.66 \pm 5.73$ & & & & & yes & no \\
9067 & 83.01 & 21018 & $23.64 \pm 3.71$ & & & & & yes & no \\
9578 & 68.36 & 23745 & $16.04 \pm 15.78$ & & & & & yes & no \\
10122 & 18.76 & 25071 & $0.78 \pm 0.29$ & & & & & yes & no \\
10122 & 18.76 & 25091 & $50.51 \pm 2.15$ & IXO 4 & NGC\,1042 ULX1& & NGC\,1042 ULX1 & no & no \\
10122 & 18.76 & 25091 & $29.54 \pm 2.40$ & IXO 4 & NGC\,1042 ULX1 & & NGC\,1042 ULX1 & no & no \\
10175 & 19.65 & 25272 & $0.90 \pm 0.20$ & & & & & yes & yes \\
10175 & 19.65 & 25272 & $0.73 \pm 0.26$ & & & & & yes & no \\
10266 & 14.57 & 25719 & $1.08 \pm 0.07$ & & & & & yes & yes \\
10266 & 14.57 & 25703 & $0.91 \pm 0.12$ & & NGC\,1068 ULX2 & & & no & yes \\
10266 & 14.57 & 25703 & $0.86 \pm 0.13$ & & NGC\,1068 ULX2 & & & no & yes \\
10314 & 9.1 & 25840 & $1.02 \pm 0.13$ & & & & NGC\,1058 ULX1 & no & yes \\
10959 & 69.36 & 26594 & $34.03 \pm 7.33$ & & & & & yes & no \\
12209 & 8.6 & 28959 & $2.81 \pm 0.11$ & IXO 6 & NGC\,1291 ULX1 & 12 & NGC\,1291 ULX3& no & yes \\
12286 & 3.70 & 29101 & $1.08 \pm 0.06$ & & & & & yes & yes \\
12286 & 3.70 & 29250 & $5.84 \pm 0.06$ & IXO 7 & NGC\,1313 ULX1 & & NGC\,1313 ULX1 & no & yes \\
12286 & 3.70 & 29250 & $11.52 \pm 0.15$ & IXO 7 & NGC\,1313 ULX1 & & NGC\,1313 ULX1 & no & yes \\
12286 & 3.70 & 29250 & $8.89 \pm 0.16$ & IXO 7 & NGC\,1313 ULX1 & & NGC\,1313 ULX1 & no & yes \\
12286 & 3.70 & 29250 & $6.61 \pm 0.13$ & IXO 7 & NGC\,1313 ULX1 & & NGC\,1313 ULX1 & no & yes \\
12286 & 3.70 & 29250 & $7.14 \pm 0.14$ & IXO 7 & NGC\,1313 ULX1 & & NGC\,1313 ULX1 & no & yes \\
12286 & 3.70 & 29250 & $11.03 \pm 0.21$ & IXO 7 & NGC\,1313 ULX1 & & NGC\,1313 ULX1 & no & yes \\
12286 & 3.70 & 29250 & $8.88 \pm 0.19$ & IXO 7 & NGC\,1313 ULX1 & & NGC\,1313 ULX1 & no & yes \\
12286 & 3.70 & 29250 & $7.98 \pm 0.23$ & IXO 7 & NGC\,1313 ULX1 & & NGC\,1313 ULX1 & no & yes \\
12286 & 3.70 & 29250 & $10.71 \pm 0.29$ & IXO 7 & NGC\,1313 ULX1 & & NGC\,1313 ULX1 & no & yes \\
12286 & 3.70 & 29250 & $7.71 \pm 0.28$ & IXO 7 & NGC\,1313 ULX1 & & NGC\,1313 ULX1 & no & yes \\
12286 & 3.70 & 29250 & $9.10 \pm 0.43$ & IXO 7 & NGC\,1313 ULX1 & & NGC\,1313 ULX1 & no & yes \\
12286 & 3.70 & 29250 & $3.23 \pm 0.06$ & IXO 7 & NGC\,1313 ULX1 & & NGC\,1313 ULX1 & no & yes \\
\\[-0.3cm]
\hline
\hline
\end{tabular}
\end{center}
\label{lastpage}
\end{table}

\end{document}